\documentclass{llncs}
\usepackage{llncsdoc}

\usepackage{url}

\usepackage{amsmath,  amssymb}
\usepackage{graphicx}
\usepackage{bm}
\usepackage{xcolor}
\usepackage{tikz}
\usepackage{todonotes}
\usepackage[boxed,lined]{algorithm2e}

\usepackage{caption}
\usepackage{subcaption}
 \usepackage{comment}
 \usepackage{longtable}
\newcommand{\vect}[1]{\boldsymbol{#1}}
\DeclareMathOperator*{\round}{round}

\SetKwFor{Algo}{Algorithm}{}{end.}

\newcommand{\RR}{\mathbb{R}}

\newcommand{\ZZ}{\mathbb{Z}}
\newcommand{\QQ}{\mathbb{Q}}

\newcommand{\vx}{\boldsymbol{x}}

\newcommand{\va}{\boldsymbol{a}}

\newcommand{\D}[2]{ \ensuremath{ \frac{d #1 }{d #2 } } }

\newcommand{\cor}[1]{{\color{black} #1}}
\newcommand{\pet}[1]{{\color{black} #1}}

\renewcommand{\thetable}{S\arabic{table}}

\begin{document}

%%%
\abovedisplayskip=3pt
\belowdisplayskip=3 pt
\abovedisplayshortskip=0pt
\belowdisplayshortskip=0pt
%%%%%

\title{Qualitative dynamics of chemical reaction networks: an investigation using partial tropical equilibrations}
\author{
Aur{\'e}lien Desoeuvres \inst{1}
\and
Peter Szmolyan \inst{2}
\and
Ovidiu Radulescu \inst{1}
}
\institute{
LPHI UMR CNRS 5235, University of Montpellier, Montpellier, France;\\
\email{ovidiu.radulescu@umontpellier.fr}
\and
Technische Universität Wien, Institute for Analysis and Scientific Computing, Vienna, Austria; 
}
\maketitle

\sloppy

\begin{abstract}
We discuss a method to describe the qualitative dynamics of chemical reaction networks
in terms of symbolic dynamics. 
The method, that can be applied to 
mass-action reaction networks with separated timescales, 
uses solutions of the 
{\em partial tropical equilibration problem} as proxies for 
symbolic states.
\cor{The partial 
tropical equilibration solutions are found algorithmically. These solutions also provide the scaling needed for slow-fast decomposition and model reduction. Any trace
of the model can thus be represented as a sequence of local approximations of the full model. We illustrate the method using as case study a biochemical model of the cell cycle. }
\end{abstract}

\section{Introduction}
Chemical reaction networks (CRN) are models of normal cell physiology and of disease and have
multiple applications in biology and medicine. Rather generally, CRNs can be described as systems 
of polynomial differential equations that result from the mass action kinetics. 
In applications one would like to characterize these models in terms of attractors, 
their bifurcations, attraction basins, and of the sequence of states 
to and on these attractors. 
These questions
belong to the qualitative theory of dynamical systems and are notoriously difficult. 

In this paper we %illustrate 
\pet{introduce} a  method
to describe the qualitative dynamics of 
mass-action law CRNs, in situations \pet{when the dynamics involves processes on 
several well separated timescales}
%the timescales of these dynamical systems 
%are well separated. 
We have suggested that in these situations, the phase space of the CRN is 
patched with slow manifolds connected to each other by continuous or discontinuous transitions \cite{Gorban2007a,Radulescu2015,Samal2016,Samal2019}. Thus,
the system \pet{stays repeatedly for  relatively long time in some metastable state before switching to some other metastable state. }
In  \cite{Radulescu2015,Samal2016} we proposed to use 
tropical equilibrations as proxies for metastable states and finite-state 
machines as discrete abstractions for the ODE dynamics of the CRN.

The concept of {\em tropical equilibration} comes from algebraic geometry and it is a necessary condition for the existence of real Puiseux series solutions of systems of polynomial equations whose coefficients are powers
or Puiseux series of some scaling parameter $\epsilon$.
The \pet{concept} is naturally \pet{related} to the \pet{problem of finding} scalings of differential equations needed in 
the {\em mathematical theory of singular perturbations} for systems with multiple timescales \cite{kruff_algorithmic_2021}.  

By revisiting the tropical scaling methodology we realized that the concept of
\pet{\em partial tropical equilibrations} is better suited to \pet{slow-fast} decompositions
than total tropical equilibrations \cite{Desoeuvres2021thesis}. The total tropical equilibration condition
means that on slow manifolds each polynomial ODE has two dominant monomial terms of opposite signs that can equilibrate each other; the flow generated by 
the remaining, un-equilibrated monomial terms is slow. However, in \pet{slow-fast decompositions} only fast variables need to be equilibrated; the
\pet{dynamics of slow variables} is governed by ODEs that may have only one dominant, but slow, 
monomial term. This leads to the concept of partial tropical equilibration that we investigate here. 

We provide an automatic method to compute partial tropical equilibrations, derived from the similar method for total tropical equilibrations based on SMT solvers, SMTcut \cite{luders2020computing}, \url{https://gitlab.com/cxxl/smtcut/-/tree/master/smtcut}. 
\cor{Our code is available at \url{https://github.com/Glawal/smtcutpartial}.}

\pet{As a case study we discuss} a six variables biochemical network describing the cyclic 
phosphorylation of different substrates in the frog embryo cell cycle, proposed by J.J.Tyson \cite{tyson1991modeling}. 
The partial equilibration
solutions of this model are grouped in branches 
that are geometrically
represented as polyhedra in the space of orders of magnitude of the 
species concentrations. The intersection relations of these polyhedra 
allow us to define a connectivity graph.
We verify numerically that 
the branches are reasonably well related to slow manifolds and that the allowed
transitions from one slow manifold to another are all edges of the
connectivity graph. 
Each branch corresponds to a reduced model that can be computed using 
the tropical scaling approach. Any trace of the model can be represented symbolically
as a sequence of branches or reduced models. Also, any trace can be approximated \pet{locally}
by solutions of the corresponding reduced model. \pet{We expect
that the global validity of these patched together local approximations
as approximations of solutions of the full model
can be shown rigorously; this is the subject of ongoing 
work.}

\section{Definitions and methods}
\subsection{Tropical geometry concepts}\label{sec:tropical}
We briefly recall here how we relate tropical geometry and singular perturbations 
using what we call tropical scaling. We follow notations from \cite{maclagan2009introduction}. The reference \cite{maclagan2009introduction} can be used by the
reader as a good introduction to tropical geometry.

We consider  differential equations whose r.h.s. are multivariate polynomials 
$f = \sum c_u x^u$, where $u$ are multi-indices
and $c_u$ coefficients. \pet{Here}, $c_u$ \pet{are considered 
to be} functions (rational powers or more generally, Puiseux series) of a  positive
scaling parameter $\epsilon$. 
We define the valuation of $c_u$ as the limit
\begin{equation}
val(c_u) = \lim_{\epsilon \to 0} \log(c_u)/\log(\epsilon).
\end{equation}
Another way to introduce valuations is via Puiseux series, i.e. power series 
with negative and positive exponents. If $c_u$ is a Puiseux series of $\epsilon$, then
$c_u \sim \epsilon^{val(c_u)}$ at the lowest order. 
As $\epsilon^{val(c_u))}$ is the dominant
term of $c_u$, the valuation of $c_u$ \cor{can be obtained from the} order of magnitude of $c_u$ \cor{at some fixed $\epsilon=\epsilon_*$}.
For $\epsilon_* = 1/10$, valuations are \cor{obtained from} decimal orders. 
With this in mind, $val(c_u))$ can  
always be found from the numerical values of the coefficients (see Section~\ref{sec:case}). 
\cor{The valuations of $x$ are unknown, so they should result from a calculation. }
The rest of this subsection is about 
the constraints on the valuations of the 
variables $x$, when $x$ satisfies polynomial 
equations. 

Given a polynomial $f = \sum c_u x^u$, its {\em tropicalization} $trop(f)$
is the piecewise-linear function 
\begin{equation}\label{eq:trop}
trop(f) (w) = min(val(c_u) + u \cdot w).
\end{equation}

The variety $V(f)$ is the set of all $x$ solutions of $f(x)=0$. 
The {\em tropical hypersurface} $trop(V(f))$ is the set of $w$ where the
minimum in $trop(f)$ is attained at least twice. A theorem of 
Kapranov relates the tropical hypersurface to the set of all possible valuations 
of $x$ on $V(f)$, namely $trop(V(f))$ is the closure of $val(x)$
where $x\in V(f)$ \cite{maclagan2009introduction}. In short, if we know the orders of $c_u$, the
orders of $x$ satisfying $f(x)=0$ are given by $trop(V(f))$.

Kapranov's theorem refers to solutions of $f=0$ in the complex field. If we are interested in
the valuations of real positive solutions of $f=0$, where $c_u$ are all real,
then one has to consider
{\em tropical equilibrations}. A tropical equilibration is 
a $w$ where the minimum in $trop(f)$ is attained at least twice, for 
at least one positive and at least one negative monomial \cite{Noel2012,Noel2014,Radulescu2015a}.  
The tropical equilibrations are thus 
possible valuations of the real positive solutions 
of $f(x) = 0$.

If $f=(f_1,\ldots,f_n)$ is a polynomial vector field,
we define $$trop(f)(w_1,\ldots,w_n)=(trop(f_1)(w_1),\ldots,trop(f_n)(w_n)).$$
The valuations
$val(x) = (val(x_1),\ldots,val(x_n))$
of the solutions of $f(x)=0$ are in the intersection of the tropical hypersurfaces of 
the component \mbox{polynomial $f_i$.} This intersection is called {\em tropical prevariety}. 
By ``abus de langage'' we call tropical equilibration also an element of the tropical prevariety that is a tropical equilibration for each component. 
%A tropical equilibration of $f$ is
%a $w = val(x)$ such that 

\subsection{Partial tropical equilibrations and \pet{slow-fast decompositions} }\label{sec:partial}
\pet{In the mathematical theory of slow-fast systems it is usually
assumed that the governing equations have the form 
\begin{eqnarray}
\D{x}{t} &=& f(x,y), \notag \\
\D{y}{t} &=& \epsilon g(x,y), \label{eq:fsmath}
\end{eqnarray}
i.e. the variables are a priori split into
the fast variable $x$ and the slow variable $y$; here 
$0 < \epsilon \ll 1$ is a suitable parameter measuring time-scale separation.
The CRNs we have in mind are typically not given in this form which is a major obstacle in using slow-fast decompositions in their analysis.
We will use partial tropical equilibrations to overcome these difficulties.}

\pet{ We start with an arbitrary splitting of the variables into 
 two groups denoted by $x$ and $y$ 
 \begin{eqnarray}
\D{x}{t} &=& f(x,y;k), \notag \\
\D{y}{t} &=& g(x,y;k), \label{eq:fs}
\end{eqnarray}
where $f$ and $g$ are polynomial vector fields whose coefficients include the kinetic parameters $k$.
Now we search for scalings and conditions leading to time scale separation similar to Eq. \ref{eq:fsmath}.
To this purpose the model under study is considered to belong to a family of models indexed by $\epsilon$.}
More precisely, the \cor{kinetic parameters $k$ are considered to be powers of $\epsilon$, $k(\epsilon) \sim \epsilon^\gamma$ (this implies
that the coefficients of $f,g$ are} 
 Puiseux series of $\epsilon$). Then, we consider the solutions of models from
this family in the limit $\epsilon \to 0$. 
\cor{The studied model is just a member of the family, obtained for a particular value $\epsilon_*$
of $\epsilon$ and having kinetic parameters $k(\epsilon_*)$.}
If the value $\epsilon_*$ placing the studied model
in the family is small enough, one may expect that the limit solution is a good approximation
for the model's solution. 

The tropicalization is useful in the scaling process, because it allows to compute the lowest order
terms of the Puiseux series expansions of the polynomials $f$ and $g$. We have
\begin{eqnarray}
\D{\bar x}{t} &=& \epsilon^{\mu_x} \bar f(\bar x, \bar y; \cor{\bar k}) + \text{ higher order terms, }\notag \\
\D{\bar y}{t} &=& \epsilon^{\mu_y} \bar g(\bar x, \bar y; \cor{\bar k}) + \text{ higher order terms, } \label{eq:fss}
\end{eqnarray}
where $\mu_x = trop(f)(val(x),val(y))  - val(x)$,  $\mu_y = trop(g)(val(x),val(y))  - val(y)$, and $\bar x, \bar y, \bar f, \bar g, \cor{\bar k}$ have
valuation zero.
$\mu_x$ and $\mu_y$ are the timescale orders (in fact orders of reciprocal timescales) for the variations of $x$ and $y$, respectively; smaller timescale orders mean faster variables. 

$\bar f$ and $\bar g$ are called tropically truncated versions of $f$ and $g$. 

Let us denote by $\mu_x \lhd \mu_y$ the set of inequalities 
$\{ (\mu_x)_i < (\mu_y)_j\,  \forall i,j\}$, meaning  
that variables
$x$ are faster than variables $y$.

If $\mu_x \lhd \mu_y$ 
one shows that the solutions of \eqref{eq:fss} converge to the solutions of the following reduced system
\begin{eqnarray}
0 &=&  \bar f(\bar x, \bar y; \cor{\bar k}), \notag \\
\D{\bar y}{t} &=& \epsilon^{d_y} \bar g(\bar x, \bar y, \cor{\bar k}), \label{eq:fsr}
\end{eqnarray}
under some conditions meaning roughly that the solutions of  $\bar f(\bar x, \bar y)=0$ \pet{are hyperbolic attracting equilibria of the equation 
$$\D{\bar x}{t} = \bar f(\bar x, \bar y; \cor{\bar k}),$$}
see \cite{kruff_algorithmic_2021} for the rigorous statement. 
The first equation of \eqref{eq:fsr} defines
the quasi-steady \pet{state} variety and imposes
constraints on $x,y$. Using the tropical approach we transform these constraints into constraints
on the order of magnitudes $val(x), val(y)$.

As a matter of fact, the valuations of $x$ and $y$ are constrained by
\begin{equation}
\left\{
\begin{array}{c}
 (val(x),val(y)) \text{ is a tropical equilibration of }  f,  \\[1mm]
 trop(f)(val(x),val(y))  - val(x) \lhd trop(g)(val(x),val(y))  - val(y). 
 \end{array}
 \right. \label{partialeq}
\end{equation}
The first of the equations \eqref{partialeq} follows from Kapranov's theorem 
because $x$ satisfies $f(x,y)=0$. The second equation is simply a condition
on the timescales. 

We call any solution of \eqref{partialeq} {\em partial tropical equilibration}.
Geometrically, \eqref{partialeq}
defines polyhedral complexes in the space of
valuations. 

If the system \eqref{eq:fs} has conservation laws, i.e. linear or polynomial 
functions $c(x,y)$ such that $\frac{\partial{c}}{\partial{x}}f + \frac{\partial{c}}{\partial{y}}g=0$ identically, one needs to 
consider the quasi-state \pet{state} equation
$f(x,y)=0$ together with the conservation equation
$c(x,y) - c_0 = 0$ where $c_0$ is constant. In this
case 
the problem \eqref{partialeq} becomes
\begin{equation}
\left\{
\begin{array}{c}
 (val(x),val(y)) \text{ is a tropical equilibration of }  f  \text{ and of }   c(x,y) - c_0,\\[1mm]
 trop(f)(val(x),val(y))  - val(x) \lhd trop(g)(val(x),val(y))  - val(y). 
 \end{array}
 \right. \label{partialeqc}
\end{equation}

We call {\em tropical scaling} of a polynomial ODE system, a fixed choice of the valuations of the polynomial coefficients and variables satisfying the partial
tropical equilibration constraints. \pet{It is very important to keep in mind,}
that different scalings will be valid in different regions of the phase space and will lead to different reductions. 

Like total tropical equilibrations, partial tropical 
equilibrations \eqref{partialeq} 
can be grouped into branches \cite{Samal2015a,samal_geometric_2016}. In a branch, the tropically truncated functions $\bar f$ and $\bar g$
are fixed. In other words, the dominant terms, 
corresponding to the $min$ value in $trop(f)$ and $trop(g)$ are the same for all solutions in a branch. Geometrically, a branch is a polyhedral face of the
polyhedral complex. 

\cor{\subsection{Coarse graining}
The model has continuous parameters and variables \pet{which from the mathematical point of view can vary in $[0,\infty)$}. Therefore, the valuations and in consequence the scales are in principle continuous. %However, in singular perturbation theory 
\pet{However, to obtain a finite number of useful approximating systems one has to use a suitable selection of discrete (often integer) scales, that cover the relevant domains in parameter- and phase space. }

In order to do so, we use {\em logarithmic paper coarse graining}. The space of parameters
and the phase space are discretized in such a way that any two parameters or two variable values taken from the
same cell of a discretizing mesh grid
have the same image on logarithmic paper. 

In this approach, positive real quantities $x$ are mapped to the
logarithmic paper
using the application
\begin{equation}\label{eq:rounding}
x  \to \frac{1}{d} round (\frac{\log x}{\log \epsilon_*} d), 
\end{equation}
where $round$ is the rounding to the nearest integer,
$\epsilon_*$ is a fixed value of $\epsilon$, $d$ is an integer,
and $x$ is any 
positive quantity, \pet{e.g. a} kinetic parameter, concentration, monomial or polynomial of concentrations. The image of $x$ via the mapping
\eqref{eq:rounding} represents the order of magnitude of $x$, which is
an integer for $d=1$ and a rational number from $\ZZ/d$ when $d>1$.

Using \eqref{eq:rounding} two values $x_1$ and $x_2$ have different images on logarithmic paper if %and only if
\begin{equation}\label{eq:grain}
| \log (x_1) - \log (x_2) | > \frac{|\log \epsilon_*|}{d}. 
\end{equation}
Equation \eqref{eq:grain} specifies the cell-size on logarithmic paper. 
For a given $\epsilon_*$,  the largest cell-size is obtained for 
integer values $d=1$. The cell-size increases 
when $\epsilon_*$ decreases. The limit $d\to \infty$ corresponds to 
the continuum. 

In practice, we want to choose an intermediate cell-size. This should not be too
small, to avoid continuous scaling, and not too large, to avoid loss of the 
structure. 
Although two parameters are in play, one can not change them independently. 
As a matter of fact, the value of $\epsilon_*$ is dictated by singular perturbations; we want this to be small enough. Therefore, the cell-size adjustment is performed
by changing $d$ after the choice of $\epsilon_*$.

Given $\epsilon_*$, $d$ the optimal $d$ corresponds to the largest cell-size that distinguishes the most robust structural features (geometry of branches of tropical equil}ibration solutions, differences 
between orders of magnitude of parameters, concentrations, monomials, see also Section~\ref{sec:symbolic} and Figure~\ref{fig:cycled1e11}).

%Right now we use trial and error; empirical criteria will be formalized in future work.  

%\todo{to illustrate these concepts we need a version of figure 3 with two different values of d, two subfigures, same epsilon, pdf svp : trace001d10cyclegrid.pdf}

\section{Case study: a cell cycle model}\label{sec:case}
The model presented is a modification of the original Tyson cell cycle model \cite{tyson1991modeling}. First,
by \pet{considering} constant %species
\pet{concentrations as} parameters,
the model has been converted to an ODE system with six variables. In order to simplify the analysis %of this model
we have \pet{also} removed a \pet{variable}
$x_6$ that does not interact with the rest of the model. This gives us the following model:
\begin{align}
\begin{split}
    \dot{x_1} = &\ k_1x_3 - k_2x_1 + k_3x_2,\\
    \dot{x_2} = &\ k_2x_1 - k_3x_2 - k_4x_2x_5,\\
    \dot{x_3} = &\ k_{10}x_4 - k_1x_3 + k_9x_3^2x_4,\\
    \dot{x_4} = &\ k_4x_2x_5 - k_9x_3^2x_4 - k_{10}x_4,\\
    \dot{x_5} = &\ k_6 - k_4x_2x_5.
\end{split}
\label{eq:cellcycle}
\end{align}
With the parameter values 
$k_1 = 1,\ 
k_2 = 1000000,\ 
k_3 = 1000,\ 
k_4 = 200,\ 
k_6 = 3/200,\ 
k_8 = 3/5,\ 
k_9 = 180,\ 
k_{10} = 9/500,\ 
k_{14} = 1,$
where $k_{14}$ is the total \pet{concentration} associated to \pet{the} conservation law $k_{14}=x_1+x_2+x_3+x_4$.

\subsection{Tropical scaling of the cell cycle model}
\label{sec:scale}
\pet{The reaction rate constants and the concentrations in this model have very different orders of magnitude which suggests dynamics on many well separated time scales. To identify these} we rescale the model \pet{as a first step of using the 
procedure described abstractly in Sect. 2:}
\begin{itemize}
    \item Consider $0<\epsilon_*<1$.
    \item Write $k_i=\bar{k}_i\epsilon^{\gamma_i}_*$ and $x_i=\bar x_i\epsilon^{a_i}_*$; thus $\gamma_i=val(k_i)$ and $a_i = val (x_i)$.
    \item The exponents $\gamma_i$ are computed
    from the numerical values of the parameters. We use 
    $\gamma_i=\round(d\log(|k_i|)/\log(\epsilon_*))/d$ to obtain rational (integer if $d=1$) exponents. These exponents become valuations when we \pet{view the cell cycle model 
    as being part of a family} of models with the same structure and with parameters $k_i(\epsilon) = \bar{k}_i\epsilon^{\gamma_i}$ %, and we take 
    \pet{in the limit $\epsilon \to 0$.}
\item
 Using this scaling we compute  the rescaled system
    \begin{align}
    \begin{split}
        \epsilon^{a_1}\dot{\bar{x_1}} = &\ \bar{k}_1\bar x_3\epsilon^{\gamma_1+a_3} - \bar{k}_2\bar x_1\epsilon^{\gamma_2+a_1} + \bar{k}_3\bar x_2\epsilon^{\gamma_3+a_2},\\
        \epsilon^{a_2}\dot{\bar{x_2}} = &\ \bar{k}_2\bar x_1\epsilon^{\gamma_2+a_1} - \bar{k}_3\bar x_2\epsilon^{\gamma_3+a_2} - \bar{k}_4\bar x_2\bar x_5\epsilon^{\gamma_4+a_2+a_5},\\
        \epsilon^{a_3}\dot{\bar{x_3}} = &\ \bar{k}_{10}\bar x_4\epsilon^{\gamma_{10}+a_4} - \bar{k}_1\bar x_3\epsilon^{\gamma_1+a_3} + \bar{k}_9\bar x_3^2\bar x_4\epsilon^{\gamma_9+2a_3+a_4},\\
        \epsilon^{a_4}\dot{\bar{x_4}} = &\ \bar{k}_4\bar x_2\bar x_5\epsilon^{\gamma_4+a_2+a_5} - \bar{k}_9\bar x_3^2\bar x_4\epsilon^{\gamma_9+2a_3+a_4} - \bar{k}_{10}\bar x_4\epsilon^{\gamma_{10}+a_4},\\
        \epsilon^{a_5}\dot{\bar{x_5}} = &\ \bar{k}_6\epsilon^{\gamma_6} - \bar{k}_4\bar x_2\bar x_5\epsilon^{\gamma_4+a_2+a_5},
    \end{split}
    \label{eq:scaledtyson}
    \end{align}
and the timescale orders of each \pet{variable}:
\begin{align}
    \begin{split}
        \mu_1 = &\ \min(\gamma_1+a_3,\ \gamma_2+a_1,\ \gamma_3+a_2)-a_1,\\
        \mu_2 = &\ \min(\gamma_2+a_1,\ \gamma_3+a_2,\ \gamma_4+a_2+a_5)-a_2,\\
        \mu_3 = &\ \min(\gamma_{10}+a_4,\ \gamma_1+a_3,\ \gamma_9+2a_3+a_4)-a_3,\\
        \mu_4 = &\ \min(\gamma_4+a_2+a_5,\ \gamma_9+2a_3+a_4,\ \gamma_{10}+a_4)-a_4,\\
        \mu_5 = &\ \min(\gamma_6,\ \gamma_4+a_2+a_5)-a_5.
    \end{split}
\end{align}
\item
%We consider that
\pet{In this setting} \cor{the slow-fast decomposition follows from the timescale orders. Instead of renaming slow variables as $y$ like in Section~\ref{sec:partial}, we define
instead a subset $S\subset \{1,\ldots,n\}$ containing indices
of the slow components.  
Thus, all variables $x_j$ with indices $j\in S$ are slow 
and the remaining variables $x_i$}  are faster 
%than the species in $S$, 
\pet{iff}  $\mu_i < \mu_j$ for all $i \notin S$, $j \in S$.
\item
\pet{Self-consistently, %we consider that 
the valuations $a_i$ }are solutions of the partial tropical equilibration problem for a fast/slow splitting with 
slow \pet{variables} $S$ (\pet{see the following subsection 3.2} and Appendix 1).
\end{itemize}

\subsection{Calculation of the partial tropical
equilibrations}
The solutions of the partial tropical equilibration problem for $S$ form a polyhedral complex, each face encoding one combinatorial possibility in the equations. 
The partial equilibration problem for $S$ can be decomposed into two kinds of constraints: \cor{i) equilibration of fast species and conservation laws,
and ii) timescale orders constraints resulting from the slow/fast decomposition.}

For example, suppose that we are interested in the partial tropical equilibration when $S=\{x_3\}$, then the problem is given by:
\begin{align}
    \begin{split}
        \gamma_1+a_3 = &\ \min(\gamma_2+a_1,\ \gamma_3+a_2),\\
        \gamma_2+a_1 = &\ \min(\gamma_3+a_2,\ \gamma_4+a_2+a_5),\\
        %\gamma_1+a_3 = &\ \min(\gamma_{10}+a_4,\ \gamma_9+2a_3+a_4)\\
        \gamma_4+a_2+a_5 = &\ \min(\gamma_9+2a_3+a_4,\ \gamma_{10}+a_4),\\
        \gamma_6 = &\ \gamma_4+a_2+a_5,\\
        \gamma_{14} = &\ \min(a_1,\ a_2,\ a_3,\ a_4),\\
       \mu_3 > \ \mu_1,\,
        \mu_3 > &\ \mu_2,\,
        \mu_3 > \ \mu_4,\,
        \mu_3 > \ \mu_5.\\
    \end{split} 
    \label{eq:pepx3}
\end{align}
The first four equations in \eqref{eq:pepx3} come
for the tropicalization of the polynomial vector field
for the fast species $\{x_1,x_2,x_4,x_5\}$. The fifth 
equation results from the tropicalization of the 
linear conservation law $k_{14} = x_1 + x_2 + x_3 + x_4$. The remaining equations simply mean that the species $x_3$ is slower than all the other. 

\cor{The real solutions of \eqref{eq:pepx3}, $\vect{a} \in \RR^n$, form a 
polyhedral complex. Coarse graining means that we look for solutions 
of \eqref{eq:pepx3} in $(\ZZ/d)^n$, in which case we obtain a discrete
set of points in the polyhedral complex. 
}
To solve this problem we treat each constraint separately. For each constraint, each choice of minima leads to a polytope $P$. The \pet{solution}
of the problem is the intersection of unions of such polytopes. 

The first equation in \eqref{eq:pepx3} 
%$\gamma_1+a_3 = \min(\gamma_2+a_1,\ \gamma_3+a_2)$ 
leads to two possibilities that each gives a polytope. When $\min(\gamma_2+a_1,\ \gamma_3+a_2) = \gamma_2+a_1$, the polytope is given by
\begin{align}
    \begin{split}
        \gamma_1+a_3 = &\ \gamma_2+a_1,\\
        \gamma_2+a_1 \leq &\ \gamma_3+a_2,
    \end{split}
\end{align}
whereas when $\min(\gamma_2+a_1,\ \gamma_3+a_2) = \gamma_3+a_2$, the polytope is given by 
\begin{align}
    \begin{split}
        \gamma_1+a_3 = &\ \gamma_3+a_2,\\
        \gamma_3+a_2 \leq &\ \gamma_2+a_1.
    \end{split}
\end{align}
So, for this equation, we get two polytopes. We  call this union of polytopes a bag, \pet{denoted as} $B_1$.

The last equation in \eqref{eq:pepx3} 
comes from the timescale constraints and has the form:
\begin{align}
    \begin{split}
        \min(\gamma_{10}+a_4,\ \gamma_1+a_3,\ \gamma_9+2a_3+a_4)-a_3 > &\  \min(\gamma_6,\ \gamma_4+a_2+a_5)-a_5,
    \end{split}
\end{align}
which gives, if $\min(\gamma_{10}+a_4,\ \gamma_1+a_3,\ \gamma_9+2a_3+a_4)=\gamma_1+a_3$ and $\min(\gamma_6,\ \gamma_4+a_2+a_5)=\gamma_4+a_2+a_5$ the polytope given by equations:
\begin{align}
    \begin{split}
        \gamma_1 > &\ \gamma_4+a_2,\\
        \gamma_1+a_3 \leq &\ \gamma_{10}+a_4,\\
        \gamma_1+a_3 \leq &\ \gamma_9+2a_3+a_4,\\
        \gamma_4+a_2+a_5 \leq &\ \gamma_6.
    \end{split}
\end{align}

The partial tropical equilibration problem is then to compute the intersection of these bags.
Generally, the problem reads:
$$A= \bigcap_{i=1}^{n_e} B_i = \bigcap_{i=1}^{n_e} \bigcup_{j=1}^{c_i} P_{ij},$$
% = \bigcup_{j=1}^{c_i} \bigcap_{i=1}^{n_e} P_{ij}$$
where $n_e$ is the number of equations in the partial tropical equilibration problem, $c_i$ is the number of choices for the equation $i$. 
As intersections and unions are in finite numbers we can reverse them and get:
$$A = \bigcap_{i=1}^{n_e} \bigcup_{j=1}^{c} P_{ij} = \bigcup_{j=1}^{c} \bigcap_{i=1}^{n_e} P_{ij},$$
where $c$ is the total number of choices $c=\sum c_i$, and a branch, which is a face of the polyhedral complex, is given by the intersection of each polytopes for a given choice $\bigcap_{i=1}^{n_e} P_{ij}$.

%\todo{write an appendix with details of implementation of the algorithm}

For our example, given $S = \{x_3\}$, 
most of the choices lead to an empty set. 
There are thus only
 two branches (computed with $\epsilon_*=1/11, d=1$): $\{x_3\}_{00}$ and $\{x_3\}_{01}$. 

If we consider the five variables cell cycle model, there are $2^5=32$ possible partial equilibration problems. One of them will not be considered as 
it consists of no constraints, this is when $S$ is the set of all species. 

We have tested the remaining 31 possibilities: the total tropical equilibrations when $S=\emptyset$, and 30 partial equilibrations. Only 9 are non-empty:  
8 partial tropical equilibrations and the total one.

Denoting a partial tropical equilibration problem by the associated set of slow species $S$, the list of solutions 
of all partial equilibration problems
is given by 
$\Big\{\emptyset,\{x_3\},\{x_4\},\{x_5\},\{x_3,x_4\},\{x_3,x_5\},\{x_3,x_4,x_5\},\{x_1,x_3,x_4,x_5\}, \{x_2,x_3,x_4,x_5\}\Big\}.$

The solutions of each partial tropical equilibration problem are grouped in a number of branches.
Each branch is denoted by an index number, for instance $\{x_3\}=\Big\{\{x_3\}_{00},\{x_3\}_{01}\Big\}$.

Although each partial tropical equilibration is 
a polyhedral complex, their union is generally 
not a complex. As a matter of fact, the intersection of two polyhedra of different partial tropical equilibrations can be just part of a face. 
In \cite{samal_geometric_2016} we have introduced
the connectivity graph describing 
adjacency of branches as faces of the 
polyhedral complex of total tropical equilibrations:
two branches are connected if they share a face. 
We  introduce here a connectivity graph for 
partial tropical equilibrations. 
A connectivity graph is a undirected graph whose vertices
are partial tropical equilibration branches. 
The connection between branches is somehow intermediate
between adjacency and incidence. Two partial tropical equilibration branches are connected if their intersection
has maximum dimension 
(the dimension of the intersection is equal to the dimension of one of the attached polyhedra).   

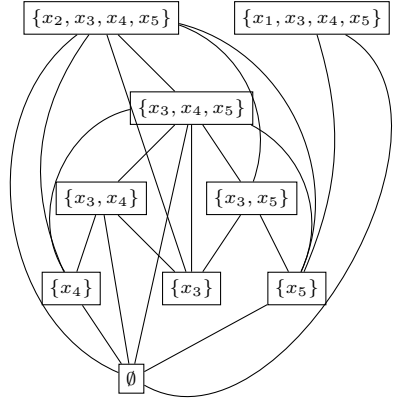
\begin{figure}
    \centering
{\scriptsize
\begin{tikzpicture}[scale=0.4]
\node[draw] (total) at (2,0) {$\emptyset$};
\node[draw] (x4) at (0,3) {$\{x_4\}$};
\node[draw] (x3) at (4,3) {$\{x_3\}$};
\node[draw] (x5) at (7.5,3) {$\{x_5\}$};
\node[draw] (x3x4) at (1,6) {$\{x_3,x_4\}$};
\node[draw] (x3x5) at (6,6) {$\{x_3,x_5\}$};
\node[draw] (x3x4x5) at (4,9) {$\{x_3,x_4,x_5\}$};
\node[draw] (x1x3x4x5) at (8,12) {$\{x_1,x_3,x_4,x_5\}$};
\node[draw] (x2x3x4x5) at (1,12) {$\{x_2,x_3,x_4,x_5\}$};
\draw (total) to[bend right=90] (x1x3x4x5);
\draw (total) to[bend left=60] (x2x3x4x5);
\draw (total) -- (x3x4);
\draw (total) -- (x3x4x5);
\draw (total) -- (x4);
\draw (total) -- (x5);
%\draw[dashed] (x1x3x4x5) -- (x2x3x4x5);
%\draw[dashed] (x3x4x5) -- (x1x3x4x5);
%\draw[dashed] (x3x5) to[bend right=20] (x1x3x4x5);
\draw (x5) to[bend right=20] (x1x3x4x5);
\draw (x3) -- (x2x3x4x5);
%\draw[dashed] (x3x4) -- (x2x3x4x5);
\draw (x3x4x5) -- (x2x3x4x5);
\draw (x4) to[bend left] (x2x3x4x5);
\draw (x5) to[bend right=50] (x2x3x4x5);
\draw (x3x5) to[bend right=45] (x2x3x4x5);
\draw (x3) -- (x3x4);
\draw (x3) -- (x3x4x5);
\draw (x3) -- (x3x5);
\draw (x3x4) -- (x3x4x5);
%\draw[dashed] (x3x4) -- (x3x5);
\draw (x4) -- (x3x4);
\draw (x3x5) -- (x3x4x5);
\draw (x4) to[bend left=55] (x3x4x5);
\draw (x5) to[bend right=44] (x3x4x5);
\draw (x5) -- (x3x5);
\end{tikzpicture}
}
    \caption{Quotiented connectivity graph for $\epsilon=1/11, d=1$. Each node is the set of solution branches of a partial tropical equilibration problem (forming a polyhedral complex), denoted by the set of slow species. Each edge  means that the intersection of two polyhedra (one from each complex)
    has the same dimension as one of the  
    polyhedra (this does not necessarily mean that one polyhedron is included in the other). 
    %Else we denote the intersection by a dashed edge, meaning that the dimension of the intersection is strictly lower than the dimension of the polyhedral complex of lower dimension.
    }
    %only borders intersect. is it true that there is no other alternative? }
    \label{fig:partiallinks}
\end{figure}

The figure \ref{fig:partiallinks} shows the 
connectivity graph for all the partial
tropical equilibrations of the cell cycle model, quotiented 
over partial equilibration problems (all branches 
of one partial equilibration problem are gathered in one node). 

Remark : this graph have been made for $\epsilon_*=1/11$ and $d=1$. 
We found that the quotiented graph is robust and does not change for different $(\epsilon_*,d)$ despite the fact that some polytope branches may be different for two different $(\epsilon_*,d)$.
Indeed, for $\epsilon_*=1/11$, $d=1$, the total tropical equilibration polyhedral complex is a segment plus a half-line whereas for $d=1000$ it is a point like for $\epsilon_*=1/29$ and $d=1$.

\subsection{Symbolic dynamics by tropicalization}\label{sec:symbolic}
We expect that the traces of the 
system \eqref{eq:cellcycle} 
are most of the time in proximity of
partial tropical equilibration solutions. 
%and the connectivity graph 
Therefore we can use the tropical equilibrations for 
symbolic coding of these traces. 

In order to test this property, we have simulated for 200 min (using the solver ode3s of Matlab R2021b) 375 numerical traces of \eqref{eq:cellcycle} starting from the different sets of initial conditions respecting $x_1+x_2+x_3+x_4=1$.
For each point $x=(x_1,x_2,x_3,x_4,x_5)\in \RR^5$ of the numerical trace, we have 
computed a valuation $a=(a_1,a_2,a_3,a_4,a_5) \in \QQ^5$
using $a_i = round( d \log(|x_{i}^{(l)})| / \log(\varepsilon)) / d$. 
This allows to compute the time-scale order of each species for this point, but also to check if a species is equilibrated or not. With these informations, we can first determine in which partial tropical equilibration the point is (if it lives in a equilibration), and then we can obtain the truncated system, and so, the branch. 

\begin{figure}[h!]
    \centering
        \includegraphics[scale=0.3]{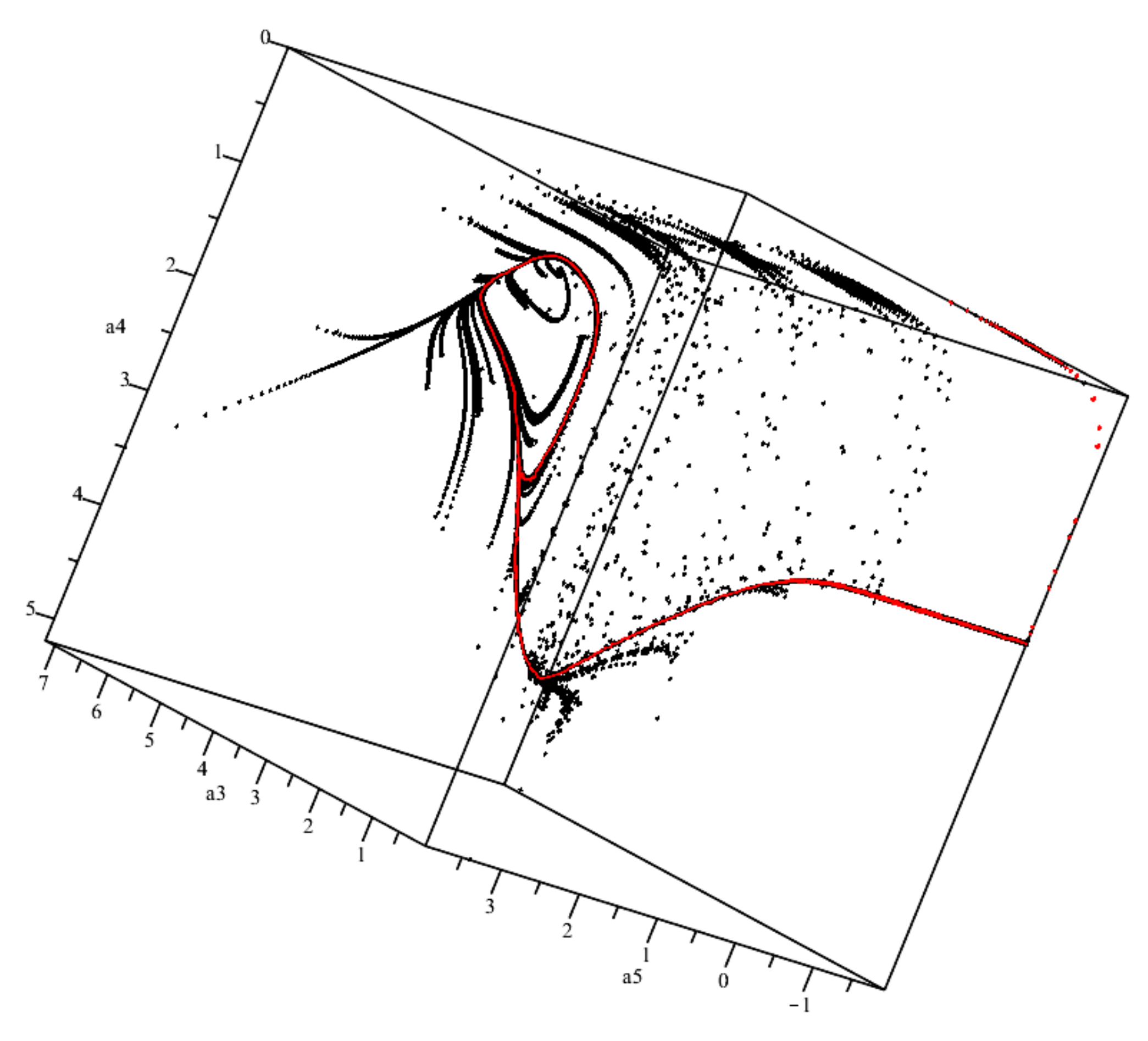}
    \caption{375 traces on the space 
    $(\log_{\epsilon_*}(x_3),\log_{\epsilon_*}(x_4),\log_{\epsilon_*}(x_5))$, $\epsilon_* = 1/11$. In red the trace starting from $(1,0,0,0,100)$.
    }
    \label{fig:alltraces3D}
\end{figure}

\begin{figure}[h!]
    \centering
    %\begin{subfigure}[b]{0.45\textwidth}
    \hskip-1truecm
  \includegraphics[scale=0.35]{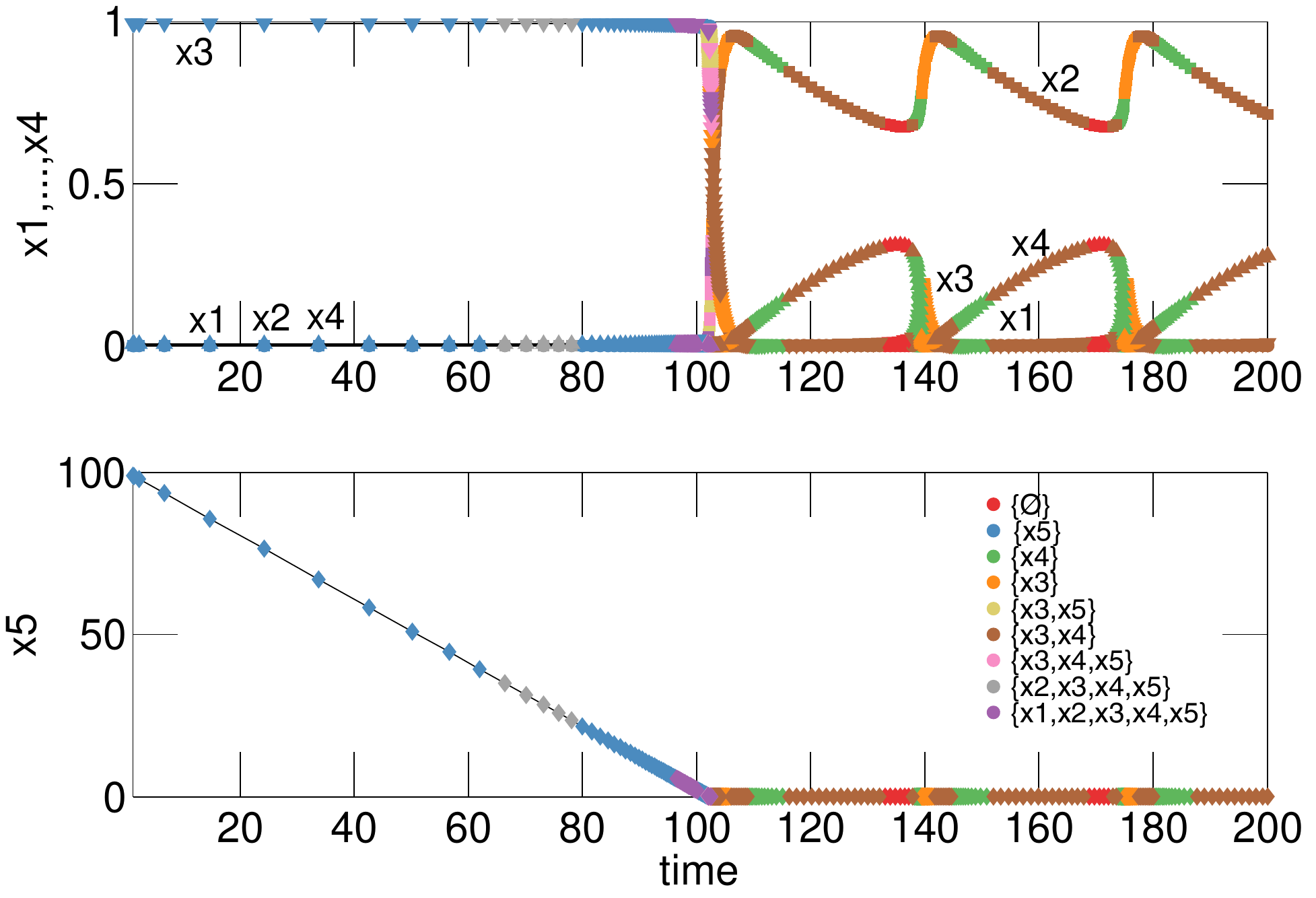}
    %    \caption{}
    %\end{subfigure}
    %\begin{subfigure}[b]{0.45\textwidth}
    %\hskip-0.5truecm
    \includegraphics[scale=0.3]{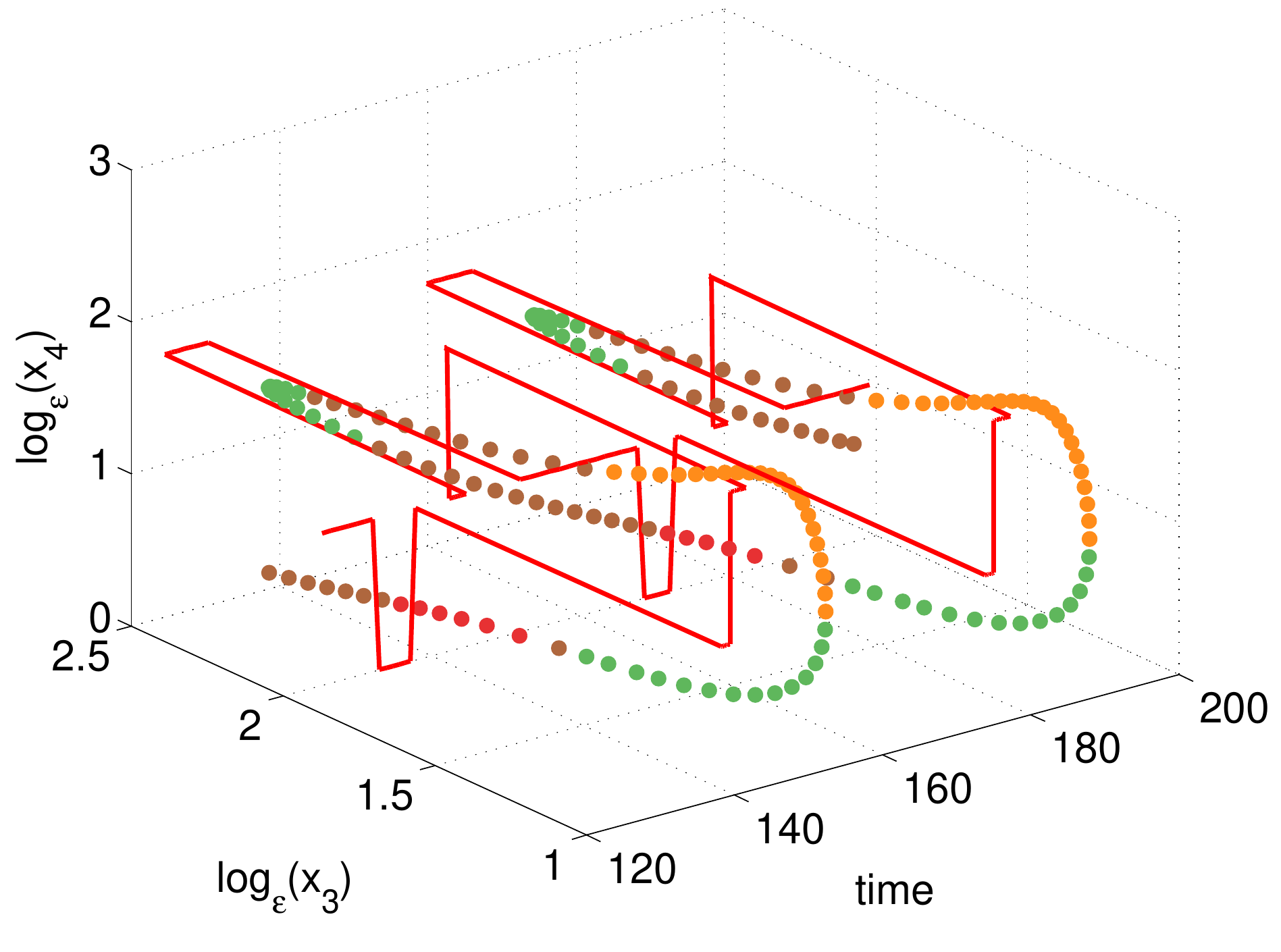}
    %\caption{}
    %\end{subfigure}
    %\includegraphics[scale=0.6]{plot3traces.pdf}
    \caption{Symbolic coding of the trace starting  at $(1,0,0,0,100)$ (a few very rapid transients states at the beginning of the trace are not represented).\cor{ \pet{On the left side}, the
    marker colors indicate the tropical equilibration branch close to which the trajectory point \pet{lies}. 
    \pet{On the right side}, the tropicalization (red line) is represented together with the continuous trace on the limit cycle part (for $\epsilon^*=1/11, d=1$).}
    %The code provides the set $S$ and indicates which species is slow in the partial tropical equilibration. For instance the code
    %$02345$ means that $S=\{x_2,x_3,x_4,x_5\}$. 
    }
    \label{fig:tracesandcoding}
\end{figure}

Making a projection of these traces on the space 
$(\log_{\epsilon_*}(x_3)$, $\log_{\epsilon_*}(x_4)$, $\log_{\epsilon_*}(x_5))$ we obtain Figure~\ref{fig:alltraces3D}. On this figure, we can see that trajectories are first converging to low dimension manifolds (dimension two or one in projection) that lead to a limit cycle.
In Figure \ref{fig:tracesandcoding}
 we have symbolically coded the points of a particularly long trajectory (marked in red), using the method described above. This figure shows that the trace follows 
 constrained transitions guided by some partial tropical equilibrations.
In order to obtain more insight, in Figure~\ref{fig:cycled1e11}
we projected the traces on the plane $(\log_{\epsilon_*}(x_3), \log_{\epsilon_*}(x_4))$.
We used this ``logarithmic paper'' representation to show also
the tropical equilibration solutions and their polyhedral branches. A tropical equilibration \cor{$a=(a_1,a_2,a_3,a_4,a_5) \in (\ZZ/d)^5$} is 
represented as a point of coordinates $(a_1,a_2)$ in this 
representation. Partial tropical equilibration branches are line segments or colored polygonal domains containing tropical equilibration points. 
\cor{As shown in Figure~\ref{fig:cycled1e11}b), a large value of $d$ ensures a small cell-size on the logarithmic paper. This
ensures a precise representation of the limit cycle, but with
a scaling that varies almost continuously and with new polygonal domains. As we are interested only in the robust features of the tropical solutions, we favor the 
value $d=1$ corresponding to   Figure~\ref{fig:cycled1e11}a).
}
% \begin{figure}
%     \centering
%     \begin{subfigure}[b]{0.4\textwidth}
%     \hskip-2truecm
%     \includegraphics[scale=0.4]{bigtrace.png}
%     \caption{}
%     \end{subfigure}
%     \begin{subfigure}[b]{0.4\textwidth}
%     \includegraphics[scale=0.35]{traceandbranches.pdf}
%     \caption{}
%     \end{subfigure}
%     \caption{a) All traces projected b) In little black dots is the trace starting at $(1,0,0,0,100)$ (alpha). Each red dot represent the trace after the transformation as in the first paragraph with $\epsilon=1/11, d=1$. Each colored region represent one branch of $\{x_3,x_4\}$. The connected union of the red and the blue lines represents $\{x_4\}$. The isolated red line represent $\{x_3\}$. The green points in $(0,2)$ and $(2,0)$ represent $\{x_5\}$. Moreover, the point $(2,0)$ represent the total equilibration $\emptyset$.}
%     \label{fig:bigtrace}
% \end{figure}
\subsection{Model reduction from partial tropical
equilibrations}
Each partial tropical equilibration solution provides a scaling of the variables.
This scaling is used for identification of slow and fast variables and for automatic model reduction with algorithms introduced in \cite{kruff_algorithmic_2021,Desoeuvres2021thesis}. The algorithms from 
\cite{kruff_algorithmic_2021} work when the quasi-steady state equations of fast 
variables satisfy hyperbolicity conditions. Very often, the hyperbolicity conditions
fail because of the existence of exact or approximate conservation laws, which 
are conservation laws of the full and tropically truncated systems, respectively. 
We showed in \cite{Desoeuvres2021thesis} that systems with full or approximate
conservation laws can be transformed into systems without conservation laws
by a change of variables. This extends the applicability of reduction 
algorithms from \cite{kruff_algorithmic_2021} to the case when there 
are conservation laws. We illustrate these techniques on an example and
refer the reader to \cite{kruff_algorithmic_2021,Desoeuvres2021thesis} for 
the complete algorithmic solutions.

Consider the scaling provided by the partial tropical equilibration solution
$a=(5,     2,     0,     2,     0)$, \cor{computed with $\epsilon_*=1/11, d=1$}. The rescaled tropically truncated 
system obtained from the original system of equations
\eqref{eq:cellcycle} reads
\begin{equation}
\begin{array}{l}
    \dot{\bar x}_{1}=\epsilon^{-6}(\bar k_3\bar x_2-\bar k_2\bar x_1),\\
    \dot{\bar x}_{2}=\epsilon^{-3}(\bar k_2\bar x_1-\bar k_3\bar x_2),\\
    \dot{\bar x}_{3}=\epsilon^{0}(\bar k_9\bar x_3^2\bar x_4-\bar k_1\bar x_3),\\
    \dot{\bar x}_{4}=\epsilon^{-2}(\bar k_4\bar x_2\bar x_5-\bar k_9\bar x_3^2\bar x_4),\\
    \dot{\bar x}_{5}=\epsilon^{0}(-\bar k_4\bar x_2\bar x_5),
\end{array}
\end{equation}
where the variables $\bar x_1$,$\bar x_2$, and $\bar x_4$ are fast.

The quasi-steady state approximation can not be applied here because the 
truncated nonrescaled system describing the fast dynamics
\begin{equation}
\begin{array}{l}
    \dot{x}_{1}=( k_3x_2- k_2x_1),\\
    \dot{x}_{2}=( k_2x_1- k_3x_2),\\
    \dot{x}_{4}=( k_4x_2x_5- k_9x_3^2x_4),
\end{array}
\end{equation}
has a conservation law $x_1 + x_2$ which is an approximate conservation law
of the full system \eqref{eq:cellcycle}.

\begin{figure}
    \centering
    \begin{subfigure}[b]{0.45\textwidth}
    \hskip-0.5truecm
        \includegraphics[scale=0.3]{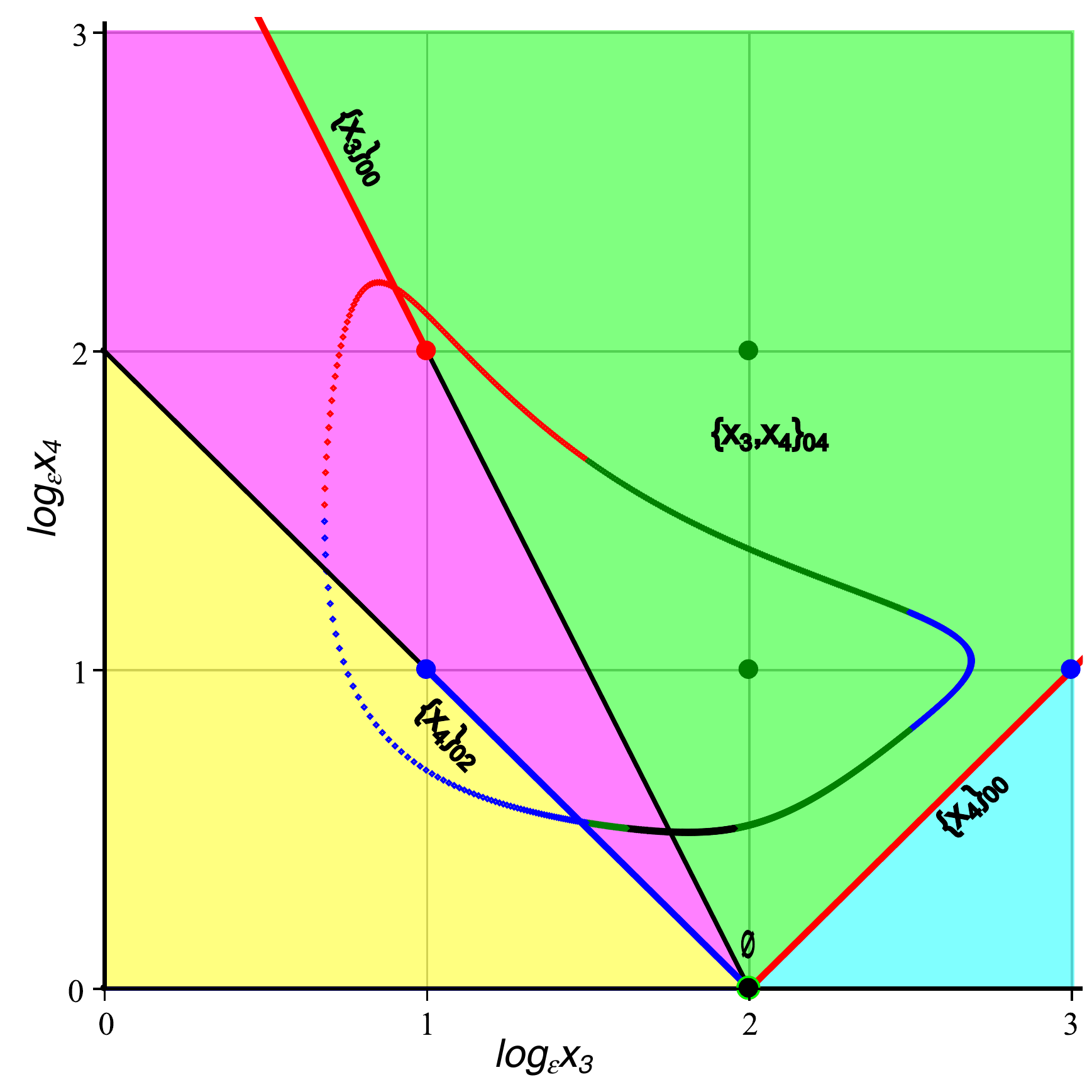}
        \caption{}
    \end{subfigure}
    \begin{subfigure}[b]{0.45\textwidth}
       % \hskip-0.5truecm
        \includegraphics[scale=0.3]{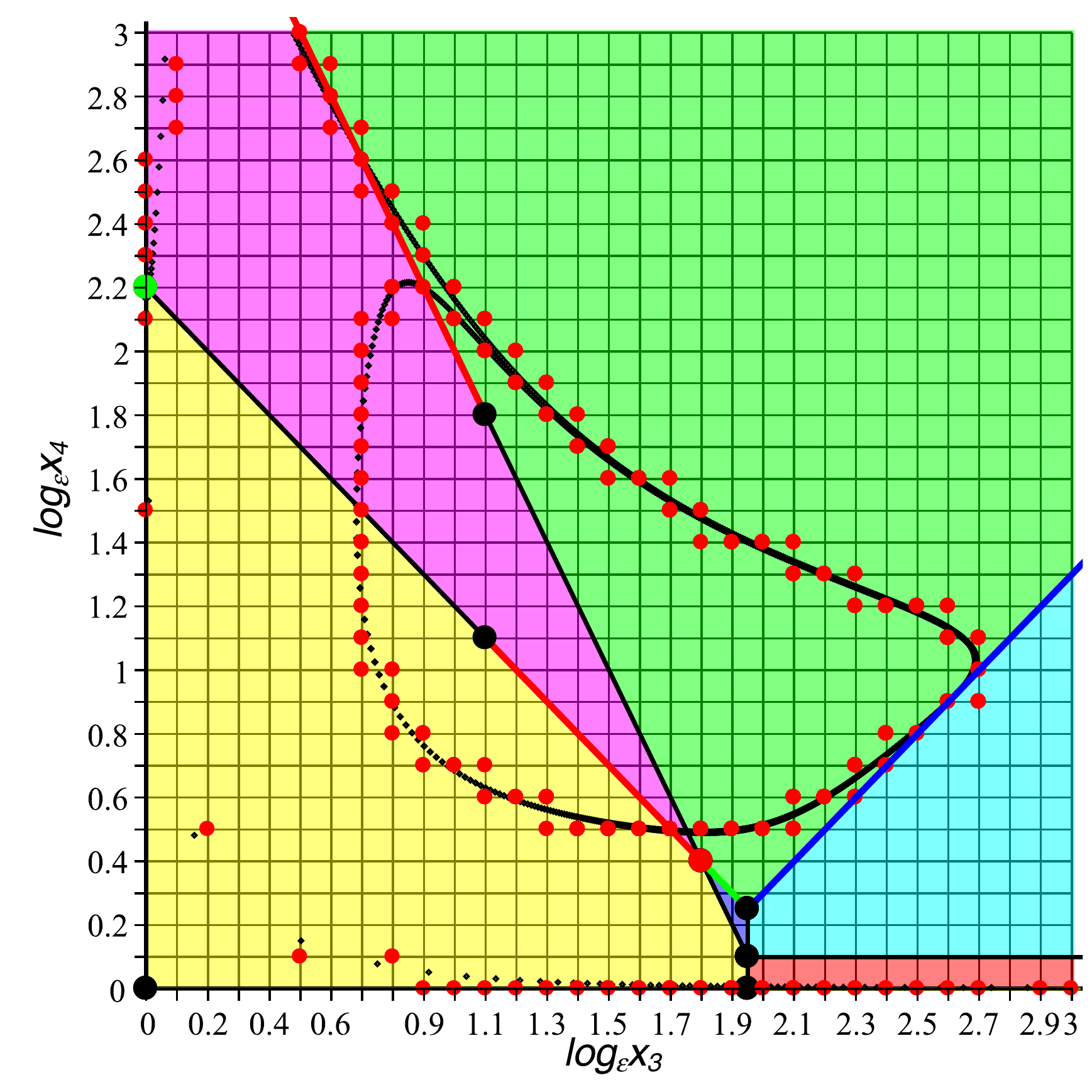}
        \caption{}
    \end{subfigure}
    \caption{a) Tropicalizing each point of the trace for $\epsilon_*=1/11$, $d=1$, we obtain six points with integer components closest to the limit cycle, each of them corresponding to a partial tropical equilibration:
    $(0,0)$ black, $(2,1), \, (2,2)$ green, $(1,1), \, (3,1)$
    blue, and $(1,2)$ red.
    We use the same color for the points of the trace
    to indicate which of the 6 points is the closest one. Black lines are intersection of two polytopes.
    %In red, the points belong to $\{x_3,x_4\}$, in blue to $\{x_4\}$, in green to $\{x_3\}$ and in black to $\emptyset$.  
    %The cycle have the following transitions : $\{x_4\}_{00}(324)\rightarrow \{x_3,x_4\}_{04}(835)\rightarrow \{\emptyset\}_{00}(249)\rightarrow \{x_3,x_4\}_{04}(31)\rightarrow \{x_4\}_{02}(73)\rightarrow \{x_3\}_{00}(116)\rightarrow \{x_3,x_4\}_{04}(148)$.
    %The point (2,0) represents the total equilibration $\emptyset$. Segments $\underbrace{((1,1),(2,0))}_{\{x_4\}_{02}}\cup\underbrace{((2,0),(4,2))}_{\{x_4\}_{00}}$ represents $\{x_4\}$. The segment $((0,4),(1,3))$ represents $\{x_3\}_{00}$. Colored regions represent $\{x_3,x_4\}$, the green one is the polyhedron $\{x_3,x_4\}_{04}$.\\
    b) For ($\epsilon_*=1/11$, $d=10$), we observe that we have more points for the limit cycle, and more polyhedra in the tropical structure. This is due to the coarse graining: 
    the grid cell is smaller for larger $d$.
    }
    \label{fig:cycled1e11}
\end{figure}

The system \eqref{eq:cellcycle} has also the exact conservation law $x_1+x_2+x_3+x_4$.

According to the method described in \cite{Desoeuvres2021thesis}, we can eliminate
all conservation laws (approximate and exact) by using the following change of
variable:
\begin{equation}
\begin{array}{l}
x_3 \leftarrow k_{14}-x_1-x_2-x_4, \\
x_2 \leftarrow x_1+x_2.
\end{array}
\end{equation}
After this change of variables we obtain a transformed system of equations that
has no conservation laws: 
\begin{equation}\label{eq:transformed}
\begin{array}{l}
\dot x_1 =       k_1k_{14} - (k_2 + k_3)x_1 - k_1x_4 + (k_3-k_1)x_2, \\
\dot x_2 =        k_1k_{14} - k_1x_2 - k_1x_4 + k_4x_1x_5 - k_4x_2x5, \\
 \dot x_4 = - (k_9k_{14}^2+ k_{10})x_4 + 2k_9k_{14}x_2x_4 + 2k_9k_{14}x_4^2 -\\ - k_9x_2^2x_4 - 2k_9x_2x_4^2 + k_4x_5x_2 - k_9x_4^3  - k_4x_1x_5, \\
 \dot x_5 = k_6 + k_4x_1x_5 - k_4x_2x_5.
 \end{array}
            \end{equation}
The rescaled tropically truncated system obtained from the transformed 
system of equations \eqref{eq:transformed} reads
\begin{equation}
\begin{array}{l}
    \dot{\bar x}_{1}=\epsilon^{-6}(\bar k_3\bar x_2-\bar k_2\bar x_1),\\
    \dot{\bar x}_{2}=\epsilon^{-2}(\bar k_1\bar k_{14}-\bar k_4\bar x_2\bar x_5),\\
    \dot{\bar x}_{4}=\epsilon^{-2}(\bar k_4\bar x_2\bar x_5-\bar k_9\bar k_{14}^2\bar x_4),\\
    \dot{\bar x}_{5}=\epsilon^{0}(-\bar k_4\bar x_2\bar x_5).
\end{array}
\end{equation}
The fast variables $x_1$, $x_2$, $x_4$ can be eliminated successively (first
$x_1$, then $x_2$ and $x_4$). One gets the reduced model that reads
 \begin{equation}
 \begin{array}{l}
     \dot x_5=-k_1 k_{14}, \\
   x_1=(k_1k_3k_{14})/(k_2k_4x_5), \\
    x_2=(k_1k_{14})/(k_4x_5),\\
    x_4=k_1/(k_9k_{14}),
 \end{array}
 \end{equation}
 in nonrescaled variables. 

The reduced model is one dimensional and describes the decrease at constant
rate of $x_5$ as can be observed in the first part of the trace shown in the 
Figure~\ref{fig:tracesandcoding}.  

\begin{figure}
    \centering
    {\scriptsize
\begin{tikzpicture}[scale=0.4]
\node[draw] (x5) at (-9,10) {$\{x_5\}:(6402\bar 2)$};
\node[draw] (x5a) at (-3,10) {$\{x_5\}:(6302\bar 1)$};
\node[draw] (x5b) at (3,10) {$\{x_5\}:(52020)$};
\node[draw] (x3x5) at (9,10) {$\{x_3,x_5\}:(41021)$};
\node[draw] (x3x5a) at (9,8) {$\{x_3,x_5\}:(41032)$};
\node[draw] (x3x4x5) at (9,6) {$\{x_3,x_4,x_5\}:(41033)$};
\node[draw] (x3x4x5a) at (9,4) {$\{x_3,x_4,x_5\}:(30044)$};
\node[draw] (x3x4) at (9,2) {$\{x_3,x_4\}:(30034)$};
\node[draw] (x3x4a) at (9,0) {$\{x_3,x_4\}:(30134)$};
\node[draw] (x3) at (3,2) {$\{x_3\}:(30124)$};
\node[draw] (x3x4b) at (-1,4) {$\{x_3,x_4\}:(30224)$};
\node[draw] (x3x4c) at (-6,2) {$\{x_3,x_4\}:(30214)$};
\node[draw] (x4) at (-6,6) {$\{x_4\}:(30314)$};
\node[draw] (total) at (-3,0) {$\emptyset$:(30204)};
\node[draw] (x4a) at (3,0) {$\{x_4\}:(30114)$};

\draw  [-to] (x5) -- (x5a);
\draw  [-to] (x5a) -- (x5b);
\draw  [-to] (x5b) -- (x3x5);
\draw  [-to] (x3x5) -- (x3x5a);
\draw  [-to] (x3x5a) -- (x3x4x5);
\draw  [-to] (x3x4x5) -- (x3x4x5a);
\draw  [-to] (x3x4x5a) -- (x3x4);
\draw  [-to] (x3x4) -- (x3x4a);
\draw  [-to] (x3x4a) -- (x3);
\draw  [-to] (x3) -- (x3x4b);
\draw  [-to] (x3x4b) -- (x3x4c);
\draw  [-to] (x3x4c) to[bend right=55] (x4);
\draw  [-to] (x4) to[bend right=55] (x3x4c);
\draw  [-to] (x3x4c) -- (total);
\draw  [-to] (total) -- (x4a);
\draw  [-to] (x4a) -- (x3);

\end{tikzpicture}
}
    \caption{Sequence of partial tropical equilibration solutions on the trace starting with (1,0,0,0,100). The branch symbol is followed by the valuation $a$. 
\pet{Here the symbol $\bar j$ means $-j$, e.g. the valuation
$(6402\bar 2)$ denotes $(6,4,0,2,-2)$.}
%    (negative components $-a_i$ are marked with $\bar a_i$).
    }
    \label{fig:symbolic_trace}
\end{figure}
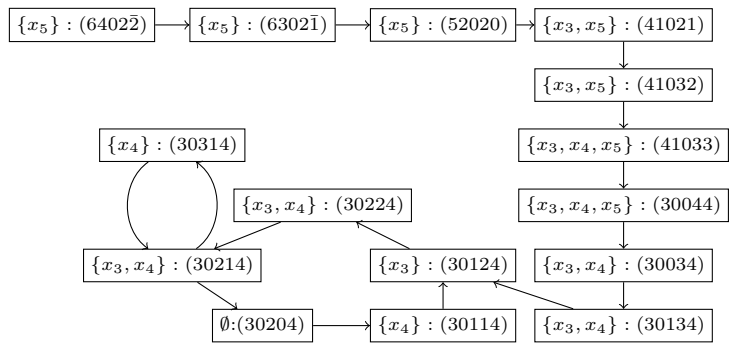

We have computed \cor{(using $\epsilon_*=1/11, d=1$)} all the reduced models for the sequence of partial tropical equilibration solutions obtained from the trace starting at $(1,0,0,0,100)$ (see Figure~\ref{fig:symbolic_trace} for a schematic representation of this sequence). The reduced models are given in the Table~\ref{tab:reduced}. 
%Several important observations can be made:

We found that scalings from the same branch lead to the same reduced model. This  
important property shows the robustness of the reduction because a branch can span several orders of magnitude of the concentrations.  

Furthermore, reduced models are nested in the sense that reduced models for scalings on a face of the polyhedral branch are supermodels (contain all the monomial terms) of reduced models originating from scalings at the interior of polyhedral branches. 

Finally, all the reduced models for solutions on the limit cycle are submodels of the 
    reduced model obtained from the total tropical equilibration that reads:
 \begin{equation}
 \begin{array}{l}
    \dot x_{2}=k_1x_3 - k_6,\\
    \dot x_{3}=k_{10}x_4 - k_1x_3 + k_9x_3^2x_4,\\
    \dot x_{4}=k_6 - k_{10}x_4 - k_9x_3^2x_4,\\
    x_1=(k_3x_2)/k_2,\\
    x_5=k_6/(k_4x_2).
 \end{array}
 \end{equation}    
    Indeed, as can be seen in Figure~\ref{fig:cycled1e11}a), the total tropical equilibration is at the intersection of all polyhedral domains corresponding to partial tropical equilibrations in the limit cycle.  This reduced model is in fact
    two dimensional ($x_3$ and $x_4$ are decoupled from $x_2$) and can be used to replace in simulation and further analysis the original five dimensional cell cycle model. 

\section{Conclusion and future work}
The tropicalization method decomposes the phase space into polyhedra within
which the dynamics of the system is simpler and can be represented by 
simpler models, with less dynamical variables and parameters. These domains
can span several orders of magnitude and therefore the resulting reduced models
are robust. 

A first possible application of this approach is to find the most 
robust reduced model, that applies to the largest domain of interest. 
For the case study discussed in this paper, we found a two variables reduced model covering the entire limit cycle and suitable for describing system's oscillatory dynamics on this attractor.

\pet{The validity of the method described in this paper
cannot be fully justified by the underlying algebraic process alone and needs to be investigated analytically
by methods from dynamical systems theory and perturbation theory. The justification of the method for small to medium size systems, e.g. the cell cycle model used in the case study, in terms of a hierarchy of slow manifolds connected by fast jumps or other transitions at states where normal hyperbolicity is lost, is the subject of ongoing work of the authors.
We expect that this will substantialy increase the understanding of  and the confidence in the method for
larger systems. }

The method also provides possible ways to approach attractors, in terms of sequences of symbolic states and local reductions. 
These abstractions can be used to model adaptation behavior of biological systems, described as
switching from one attractor to another, passing through transient states of different lifetimes. Our approch provides the timescales of each state
and the sequence of variables that are active and relax in each transient. The predicted timescales and
relaxing variables are important biologically. In the case of the cell cycle, they could 
be used to discuss  interesting dynamical regimes. In medical applications, network perturbations are
used in targeted therapies. Dynamical information is rarely taken into account when designing
such therapies. \pet{However, as shown in this paper, orders of magnitude of variables and parameters and the associated multiple timescales strongly 
influence and structure the possible dynamics and response
of the system. Thus,} together with network 
topology, \pet{orders of magnitude and }timescale information must be considered \pet{together} for predicting the effect of network perturbation.

%Blow-up method when the critical manifold loss hyperbolicity (fold points for example). Could be the case of the transition $\{x_5\}_{02}\rightarrow\{x_2,x_3,x_4,x_5\}_{11}\rightarrow\{x_5\}_{02}$.

\section*{Acknowledgements} \pet{The work of A.D. and O.R. was funded by the ANR-17-CE40-0036 project SYMBIONT and by the Campus France / BMBWF Program Amadeus 2020. The work of P.S. was partially funded by OEAD as WTZ-project FR 04/2020.}

\bibliographystyle{splncs04} 
\bibliography{radulescu,thesisAD}

\section*{Appendix 1. Calculation of partial tropical
equilibrations: algorithms and implementation. }

The input of our algorithms is a set of
 polynomial ODE system describing the CRN kinetics:
\begin{eqnarray}\label{eq:fidetail}
    \dot{x}_i &=& f_i(\vx), \, \text{where}\notag \\
    f_i(\vx) &=& \sum_{j=1}^r C_{ij} k_{j} \vx^{\vect{\alpha}_{j}} \in \RR[x_1,\dots,x_n], 1\leq i\leq n,
\end{eqnarray}
%where $r$ is the number of distinct monomials in the system, 
$C_{ij}$ are integers representing stoichiometric coefficients and $k_{j}$ are real, positive kinetic parameters.

%Let
%\begin{equation}\label{eq:lclfte}
%g_l=\sum_{j=1}^nC'_{lj} x_j =  k'_l, \quad 1 \leq l \leq n_c,
%\end{equation}
%be linear conservation laws associated to the system \eqref{eq:fidetail}.
%where $n_c$ is the number of conservation laws. 

%We transform the system in the following way, such that each real polynomial becomes a Puiseux series polynomial, allowing the use of valuation and tropicalization.

Let $0<\epsilon_*<1$ and define $\gamma_j:=\round(d\frac{\log(|k_j|)}{\log(\epsilon_*)})/d$. 

%This allow to transform the system \eqref{eq:fidetail} into a new system:
%\begin{equation}\label{eq:puiseuxsystem}
%   \tilde{f}_i(\vx) = \sum_{j=1}^r C_{ij} t^{\gamma_{j}} \vx^{\vect{\alpha}_{j}} \in %\CC\{\{t\}\}[x_1,\dots,x_n], 1\leq i\leq n
%\end{equation}
%where the powers $\gamma_j$ represent order of magnitude of the parameters $k_j$.
%We can now define the tropical equilibration as in section 2, described in detail below.

Following 
the general approach of the Sections~\ref{sec:tropical}, \ref{sec:partial}, we tropicalize the polynomials
$f_i$:
$$
Trop(f_i) =  \min_{j,C_{ij}\neq 0}(\gamma_{j}+\langle\vect{a},\vect{\alpha}_{j}\rangle).
$$

Let us consider that a $S\subset \{1,...,n\}$ subset represents the slow species of the
\eqref{eq:fidetail}.
Then, according to the definition introduced in the Section~\ref{sec:partial}, the partial tropical 
equilibration problem for $S$ consists in finding a vector $\vect{a}$ such that:
\begin{align}
    \min_{j,C_{ij}>0}(\gamma_{j}+\langle\vect{a},\vect{\alpha}_{j}\rangle) & =\min_{j,C_{ij}<0}(\gamma_{j}+\langle\vect{a},\vect{\alpha}_{j}\rangle),\ i\notin S \label{eq:ctreq}\\
    \min_{j,C'_{lj}\neq 0}(a_j) & =\gamma_l',\ 1\leq l\leq n_c \label{eq:ctrcl}\\
    \min_{j,C_{ij}\neq 0}(\gamma_{j}+\langle\vect{a},\vect{\alpha}_{j}\rangle)-a_i & <\min_{j,C_{ij}\neq 0}(\gamma_{j}+\langle\vect{a},\vect{\alpha}_{j}\rangle)-a_{i'},\ i\notin S,\ i'\in S \label{eq:ctrieq}.
\end{align}

The solution of the partial tropical equilibration for $S$ is a polyhedral complex, each maximal polyhedron (called branches and denoted $S_{index}$) being the solution of one set of choices as described below.

Let 
\begin{align}
    \begin{split}
        \rho_i & =argmin_{j,C_{ij}>0}(\gamma_{j}+\langle\vect{a},\vect{\alpha}_j\rangle)\\
        \eta_i & =argmin_{j,C_{ij}<0}(\gamma_{j}+\langle\vect{a},\vect{\alpha}_j\rangle)\\
        \zeta_i & =argmin_{j,C_{ij}\neq 0}(\gamma_{j}+\langle\vect{a},\vect{\alpha}_j\rangle - a_i)\\
        \omega_{i'} & =argmin_{j,C_{i'j}\neq 0}(\gamma_{i'j}+\langle\vect{a},\vect{\alpha}_j\rangle - a_{i'})\\
        \sigma_l & =argmin_{j,C'_{lj}\neq 0}(a_j)
    \end{split}
\end{align}

For each $i\notin S$, by fixing $\rho_i$ and $\eta_i$, we get a polyhedron consisting in the equation and inequations 
\begin{align}
\begin{split}
\gamma_{\rho_i}+\langle\vect{a},\vect{\alpha}_{\rho_i}\rangle &= \gamma_{\eta_i}+\langle\vect{a},\vect{\alpha}_{\eta_i}\rangle\\
\gamma_{\rho_i}+\langle\vect{a},\vect{\alpha}_{\rho_i}\rangle &\leq  \gamma_{j}+\langle\vect{a},\vect{\alpha}_j\rangle,\quad C_{ij}\neq 0
\end{split}
\label{eq:polforeq}
\end{align}
We have also a polyhedron for each $1\leq l\leq n_c$ by fixing $\sigma_l$, given by the equation and inequations 
\begin{align}
\begin{split}
a_{\sigma_l} &=\gamma'_l\\
a_{\sigma_l} &\leq a_j,\quad C_{lj}\neq 0.
\end{split}
\label{eq:polforeqcl}
\end{align}

For each $i\notin S, i'\in S$, by fixing $\zeta_i$ and $\omega_{i'}$, we get a polyhedron consisting in the equation and inequations 
\begin{align}
\begin{split}
\gamma_{\zeta_i}+\langle\vect{a},\vect{\alpha}_{\zeta_i}\rangle - a_i & \leq \gamma_{\omega_{i'}}+\langle\vect{a},\vect{\alpha}_{\eta_{i'}}\rangle - a_{i'}\\
\gamma_{\zeta_i}+\langle\vect{a},\vect{\alpha}_{\zeta_i}\rangle &\leq  \gamma_{j}+\langle\vect{a},\vect{\alpha}_j\rangle,\quad C_{ij}\neq 0\\
\gamma_{\omega_{i'}}+\langle\vect{a},\vect{\alpha}_{\omega_{i'}}\rangle &\leq  \gamma_{j}+\langle\vect{a},\vect{\alpha}_j\rangle,\quad C_{i'j}\neq 0\\
\end{split}
\label{eq:polforsf}
\end{align}

For each $i$ in \eqref{eq:ctreq}, by rolling the choices $(\rho_i, \eta_i)$, we get a union of polyhedra, called bag. 
For each $l$ in \eqref{eq:ctrcl}, by rolling the choices $(\sigma_l)$, we get a union of polyhedra, also called bag. 
For each $i,i'$ in \eqref{eq:ctrieq}, by rolling the choices $(\zeta_i, \omega_{i'})$, we get a union of polyhedra, also called bag. 

If we note $B_i$ these bags, the problem reads:
$$A= \bigcap_{i=1}^{n_e} B_i = \bigcap_{i=1}^{n_e} \bigcup_{j=1}^{c_i} P_{ij}$$
% = \bigcup_{j=1}^{c_i} \bigcap_{i=1}^{n_e} P_{ij}$$
where $n_e$ is the number of equations in the partial tropical equilibration problem, and $c_i$ is the number of choices for the equation $i$. 
As intersections and unions are in finite numbers we can reverse them and get:
$$A = \bigcap_{i=1}^{n_e} \bigcup_{j=1}^{c} P_{ij} = \bigcup_{j=1}^{c} \bigcap_{i=1}^{n_e} P_{ij}$$
where $c$ is the total number of choices ($c=\sum c_i$). A branch is then given by $\bigcap_{i=1}^{n_e} P_{ij}$ for one choice $j$.

To solve the tropical equilibration problem, we follow the same process as described in \cite{luders2020computing}, except that we have a different set of bags. We give here how we implement it, the code can be found at \url{https://github.com/Glawal/smtcutpartial}.

We have tested it on the model \eqref{eq:cellcycle} under python 3.7, with the solver mathsat, each solution was given under 0.5s, for a total time under 2s. The complexity of finding each tropical equilibration is exponential in the number of variable, due to the combinatorial choice of $S$. And the complexity of solving one equilibration problem is theoretically exponential in the number of bags, but the smt method used, with a preprossessing, allow to reduce this complexity by removing some choices.

As each polyhedron is given by a choice of $(\rho_i,\eta_i,\zeta_i,\omega_{i'},\sigma_l)$, we can associate to the polyhedron a tropically truncated system $E'$ defined by:

\begin{align}\label{eq:truncatedsystem}
    \begin{split}
        \bar f_i(\vx) & = C_{i\rho_i} k_{\rho_i} \vx^{\vect{\alpha}_{\rho_i}} + C_{i\eta_i} k_{\eta_i} \vx^{\vect{\alpha}_{\eta_i}},\quad i\notin S\\
        \bar f_{i'}(\vx) & = C_{i'\omega_{i'}} k_{\omega_{i'}} \vx^{\vect{\alpha}_{\omega_{i'}}}, \quad i'\in S\\
        k'_l & = C'_{l\sigma_l} x_{\sigma_l}, \quad 1 \leq l \leq n_c,
    \end{split}
\end{align}

The tropically truncated system describe the dominant dynamics of each species.

Suppose now that you have a point $\vx$ coming from a simulation. We can associate to this point a polyhedron of one of the tropical equilibration, if possible, using the following procedure. 
We compute the species concentration orders $a_i=\round(d\frac{\log(x_i)}{\log (\epsilon_*)})/d$
%. Then in equation \eqref{eq:puiseuxsystem}, we replace $\vx$ by the values $\va$ obtained 
and check if these orders are solution of one of the tropical equilibration problem \eqref{eq:ctreq}, \eqref{eq:ctrcl}, \eqref{eq:ctrieq}. To check if $\va$ is a solution, we need to compute the timescale of each species (that allow a list of $S$ choice), and to compute orders of each monomials (given by $\gamma_j+\langle \va, \vect{\alpha}_j \rangle$), which give us the list of equilibrated species. Then we choose $S$ as being minimal (each equilibrated species that are faster that non equilibrated species are considered outside $S$), this gives us the final choice $(S,\rho_i,\eta_i,\zeta_i,\omega_{i'},\sigma_l)$, denoted $S_{index}$, the index being the index of $P$ in $rr$ in Algorithm~\ref{alg:fte}.

\begin{algorithm}[htb]
\SetAlgoVlined
\Algo{{\normalsize{\tt smtcutpartial}}($\eqref{eq:fidetail}$%,$\eqref{eq:lclfte}$
,$S$)}{
$bb\gets{\tt makePolyhedraForPTE}(\eqref{eq:fidetail}%,\eqref{eq:lclfte}
,S)$\\
$rr\gets{\tt computePolyhedronDnf}(bb)$\\
\Return $rr$
}
\caption{The algorithm used to find the partial tropical equilibration for $S$.}
\label{alg:smtcutpartial}
\end{algorithm}

\begin{algorithm}
\SetAlgoVlined
\Algo{{\normalsize{\tt makePolyhedraForPTE}}($\eqref{eq:fidetail}$,$%\eqref{eq:lclfte},
S$)}{
$bb\gets \emptyset$\\
\ForEach{$i\in \{1,...,n\}\backslash S$}{
    $bb\gets bb\cup {\tt equilibrate}(\dot{x}_i)$\\
    \ForEach{$j\in S$}{
         $bb\gets bb\cup {\tt slowFastPol}(\dot{x}_i,\dot{x}_j)$   
    }
}
\Return $bb$
}
\caption{This algorithm makes the list of bags, each bag representing a list of polyhedra, such that each polyhedron is linked to an equilibration for a fast species or a conservation law, or a slow fast decomposition.}
\label{alg:ppte}
\end{algorithm}

\begin{algorithm}[htb]
\SetAlgoVlined
\Algo{{\normalsize{\tt equilibrate}}($\dot{x}_i$)}{
    $pp,np,b\gets \emptyset$\\
    \ForEach{$1\leq j \leq r_i$}{
        $t=trop(k_{j}\vx^{\vect{\alpha}_j})$\\
        \If{$S_{ij}<0$}{
            $np\gets np\cup t$
        }
        \Else{
            $pp\gets pp\cup t$
        }
    }
    \ForEach{$(a,c)\in pp\times np$}{
        $p=${\tt makePolyhedron}($a,c,pp,np$)\\
        $b\gets b\cup p$ 
    }
    \Return $b$   
}
\caption{This algorithm computes the bag $b$ linked to an equation of the system, that represent each possible equilibration. It splits negative and positive monomials, computes their tropicalization and makes the bag.}
\label{alg:equil}
\end{algorithm}

\begin{algorithm}[htb]
\SetAlgoVlined
\Algo{{\normalsize{\tt makePolyhedron}}($a,c,pp,np$)}{
$eq\gets a=c$\\
$ieq\gets \emptyset$\\
\ForEach{$i\in pp\cup np$}{
    $ieq\gets ieq\cup a\leq i$
}
$p\gets {\tt make}(eq,ieq)$\\
\Return $p$
}
\caption{$p$ is a polyhedron defined by equation \eqref{eq:polforeq} or \eqref{eq:polforeqcl}.}
\label{alg:polyhedroneq}
\end{algorithm}

\begin{algorithm}[htb]
\SetAlgoVlined
\Algo{{\normalsize{\tt slowFastPol}}($\dot{x}_i,\dot{x}_j$)}{
    $sp,fp,b\gets \emptyset$\\
    \ForEach{$1\leq m \leq r_i$}{
        $t=trop(k_{m}\vx^{\vect{\alpha}_m-a_i})$\\
        $fp\gets fp\cup t$
    }
    \ForEach{$1\leq m \leq r_j$}{
        $t=trop(k_{m}\vx^{\vect{\alpha}_m-a_j})$\\
        $sp\gets sp\cup t$
    }
    \ForEach{$(a,c)\in sp\times fp$}{
        $p=${\tt makePolyhedronSF}($a,c,sp,fp$)\\
        $b\gets b\cup p$ 
    }
    \Return $b$   
}
\caption{This algorithm computes the bag $b$ linked to a slow fast decomposition between two species $x_i$ (fast) and $x_j$ (slow). As the order the species has an impact on the slow fast decomposition, we multiply each monomial in $\dot{x}_q$ by $\frac{1}{x_q}$. Then we split each monomials, compute their tropicalization and make the bag.}
\label{alg:sfp}
\end{algorithm}

\begin{algorithm}[htb]
\SetAlgoVlined
\Algo{{\normalsize{\tt makePolyhedronSF}}($a,c,sp,fp$)}{
$eq\gets \emptyset$\\
$ieq\gets a\geq c$\\
\ForEach{$i\in sp$}{
    $ieq\gets ieq\cup a\leq i$
}
\ForEach{$i\in fp$}{
    $ieq\gets ieq\cup c\leq i$
}
$p\gets {\tt make}(eq,ieq)$\\
\Return $p$
}
\caption{$p$ is a polyhedron defined by equation \eqref{eq:polforsf}.}
\label{alg:polyhedronsf}
\end{algorithm}

\begin{algorithm}[htb]
\SetAlgoVlined
\Algo{{\normalsize{\tt computePolyhedronDnf}}($bb$)}{
$solver\gets{\tt getSolveur}(incremental=true)$\\
$f\gets{\tt convertToSMTFormula}(bb)$\\
$rr\gets \emptyset,\ bool\gets true$\\
\While{$bool$}{
    $solver.addAssertion(f)$\\
    $(x,bool)=solver.solve(f)$ /*$x$ is a point that satisfy the constraints, $bool$ is $false$ if no $x$*/\\
    \If{$Not(bool)$}{Break}
    \Else{
        $R=\emptyset$\\
        \ForEach{$b\in bb$}{
            \ForEach{$P\in b$}{
                \If{$x\in P$}{
                    $R\gets R\cup P.constraints()$\\
                    Break
                }
            }
        }
        $f\gets Not(R)$\\
        $rr\gets rr\cup R$
}
}
\Return $rr$
}
\caption{This algorithm computes the intersection of a set of bags $bb$. Each polyhedron is transformed to a logical constraint. $x$ represents a point that satisfy the set of constraints, then, if a such $x$ is found, it is contained by a polyhedron $P$ common to each equilibration. We remove this polyhedron from search by adding a constraint and continue the search until there is no feasible point. At the end we get a list of polyhedra $rr$, that is the tropical equilibration.}
\label{alg:fte}
\end{algorithm}

\clearpage
\section*{Appendix 2. Sequence of reduced models. }

\begin{center}
{\footnotesize
\begin{longtable}{|c|c|c|c|}
\caption{The sequence of reduced models corresponding to different partial tropical equilibrations along the trace starting at 
$(1,0,0,0,100)$, \cor{computed with $\epsilon_*=1/11, d=1$}. A few rapid transient states at the beginning of the trace
where not analysed and we start with
the \cor{partial tropical equilibration solution} $(6,     4,     0,     2,    -2)$.
We indicate the change of variables needed for the elimination of 
some exact and approximate conservation laws, the truncated rescaled
system indicating the local timescales of the variables and the local reduced
system in nonscaled variables and parameters. The change of variables is included in the definition of the reduced system. 
} \label{tab:reduced} \\
\hline \multicolumn{1}{|c|}{\textbf{Tropical solution}} & \multicolumn{1}{c|}{\textbf{Truncated rescaled system}} &
\multicolumn{1}{c|}{\textbf{Change of variables}} &
\multicolumn{1}{c|}{\textbf{Reduced system}} \\ \hline 
\endfirsthead

\hline 
 $\begin{array}{l}  \{x_5\}_{02} \\ (6,     4,     0,     2,    -2)\end{array}$ & 
$\begin{array}{l}
 \dot{\bar x}_1= \epsilon^{-6}(\bar k_1\bar k_{14}-\bar k_2\bar x_1)\\
    \dot{\bar x}_2= \epsilon^{-4}(\bar k_2\bar x_1-\bar k_4\bar x_2\bar x_5)\\
    \dot{\bar x}_4= \epsilon^{-2}(\bar k_4\bar x_2\bar x_5-\bar k_9\bar k_{14}^2\bar x_4)\\
    \dot{\bar x}_5= \epsilon^2(-\bar k_4\bar x_2\bar x_5)
\end{array}$  
&
$x_3 \leftarrow k_{14}-x_1-x_2-x_4$
 & 
 $\begin{array}{l}
     \dot x_5=-k_1 k_{14} \\
    x_1=(k_1k_{14})/k_2 \\
    x_2=(k_1k_{14})/(k_4x_5)\\
    x_4=k_1/(k_9k_{14})
 \end{array}$    
\\
\hline
 $\begin{array}{l}  \{x_5\}_{02} \\(6,     3,     0,     2,    -1)\end{array}$ & 
$\begin{array}{l}
    \dot{\bar x}_{1}=\epsilon^{-6}(\bar k_1\bar k_{14} + \bar k_3\bar x_2-\bar k_2\bar x_1)\\
    \dot{\bar x}_{2}=\epsilon^{-3}(\bar k_2\bar x_1- \bar k_3\bar x_2 - \bar k_4\bar x_2\bar x_5)\\
    \dot{\bar x}_{4}=\epsilon^{-2}(\bar k_4\bar x_2\bar x_5-\bar k_9\bar k_14^2\bar x_4)\\
    \dot{\bar x}_{5}=\epsilon^{1}(-\bar k_4\bar x_2\bar x_5)
\end{array}$  
&
$x_3 \leftarrow k_{14}-x_1-x_2-x_4$
 & 
 $\begin{array}{l}
    \dot x_{5}=-k_1k_{14}\\
    x_1=(k_1k_{14}(k_3 + k_4x_5))/(k_2k_4x_5)\\
    x_2=(k_1k_{14})/(k_4x_5)\\
    x_4=k_1/(k_9k_{14})
 \end{array}$    
\\
\hline
 $\begin{array}{l}  \{x_5\}_{03} \\(5,     2,     0,     2,     0)\end{array}$ & 
$\begin{array}{l}
    \dot{\bar x}_{1}=\epsilon^{-6}(\bar k_3\bar x_2-\bar k_2\bar x_1)\\
    \dot{\bar x}_{2}=\epsilon^{-2}(\bar k_1\bar k_{14}-\bar k_4\bar x_2\bar x_5)\\
    \dot{\bar x}_{4}=\epsilon^{-2}(\bar k_4\bar x_2\bar x_5-\bar k_9\bar k_{14}^2\bar x_4)\\
    \dot{\bar x}_{5}=\epsilon^{0}(-\bar k_4\bar x_2\bar x_5)
\end{array}$  
&
$\begin{array}{l}
x_3 \leftarrow k_{14}-x_1-x_2-x_4 \\
x_2 \leftarrow x_1+x_2
\end{array}$
 & 
 $\begin{array}{l}
     \dot x_5=-k_1 k_{14} \\
   x_1=(k_1k_3k_{14})/(k_2k_4x_5) \\
    x_2=(k_1k_{14})/(k_4x_5)\\
    x_4=k_1/(k_9k_{14})
 \end{array}$    
\\
\hline
 $\begin{array}{l}  \{x_3,x_5\}_{01} \\( 4,     1,     0,     2,     1)\end{array}$ & 
$\begin{array}{l}
    \dot{\bar x}_{1}=\epsilon^{-6}(\bar k_3\bar x_2-\bar k_2\bar x_1)\\
    \dot{\bar x}_{2}=\epsilon^{-1}(\bar k_1\bar k_{14}-\bar k_4\bar x_2\bar x_5)\\
    \dot{\bar x}_{4}=\epsilon^{-2}(\bar k_4\bar x_2\bar x_5-\bar k_9\bar k_{14}^2\bar x_4)\\
    \dot{\bar x}_{5}=\epsilon^{-1}(-\bar k_4\bar x_2\bar x_5)
\end{array}$  
&
$\begin{array}{l}
x_3 \leftarrow k_{14}-x_1-x_2-x_4 \\
x_2 \leftarrow x_1+x_2
\end{array}$
 & 
 $\begin{array}{l}
    \dot x_{2}=k_1k_{14} - k_4x_2x_5\\
    \dot x_{5}=-k_4x_2x_5\\
    x_1=(k_3x_2)/k_2\\
    x_4=(k_4x_2x_5)/(k_9k_{14}^2)
 \end{array}$    
\\
\hline
$\begin{array}{l} \{x_3,x_5\}_{01} \\(4,     1,     0,     3,     2)\end{array}$ & 
$\begin{array}{l}
    \dot{\bar x}_{1}=\epsilon^{-6}(\bar k_3\bar x_2-\bar k_2\bar x_1)\\
    \dot{\bar x}_{2}=\epsilon^{-1}(\bar k_1\bar k_{14})\\
    \dot{\bar x}_{4}=\epsilon^{-2}(\bar k_4\bar x_2\bar x_5-\bar k_9\bar k_{14}^2\bar x_4)\\
    \dot{\bar x}_{5}=\epsilon^{-1}(-\bar k_4\bar x_2\bar x_5)
\end{array}$  
&
$\begin{array}{l}
x_3 \leftarrow k_{14}-x_1-x_2-x_4 \\
x_2 \leftarrow x_1+x_2
\end{array}$
 & 
 $\begin{array}{l}
    \dot x_{2}=k_1k_{14}\\
    \dot x_{5}=-k_4x_2x_5\\
    x_1=(k_3x_2)/k_2\\
    x_4=(k_4x_2x_5)/(k_9k_{14}^2)
 \end{array}$    
\\
\hline
$\begin{array}{l}  \{x_3,x_4,x_5\}_{22} \\(4,     1,     0,     3,     3)\end{array}$ & 
$\begin{array}{l}
    \dot{\bar x}_{1}=\epsilon^{-6}(\bar k_3\bar x_2-\bar k_2\bar x_1)\\
    \dot{\bar x}_{2}=\epsilon^{-1}(\bar k_1\bar k_{14})\\
    \dot{\bar x}_{4}=\epsilon^{-2}(-\bar k_9\bar k_{14}^2\bar x_4)\\
    \dot{\bar x}_{5}=\epsilon^{-1}(\bar k_6-\bar k_4\bar x_2\bar x_5)
\end{array}$  
&
$\begin{array}{l}
x_3 \leftarrow k_{14}-x_1-x_2-x_4 \\
x_2 \leftarrow x_1+x_2
\end{array}$
 & 
 $\begin{array}{l}
    \dot x_{2}=k_1k_{14}\\
    \dot x_{4}=-k_9k_{14}^2x_4\\
    \dot x_{5}=k_6 - k_4x_2x_5\\
    x_1=(k_3x_2)/k_2
 \end{array}$    
\\
\hline
 $\begin{array}{l}  \{x_3\}_{00} \\(3,     0,     0,     4,     4)\end{array}$ & 
$\begin{array}{l}
    \dot{\bar x}_{1}=\epsilon^{-6}(\bar k_3\bar x_2-\bar k_2\bar x_1)\\
    \dot{\bar x}_{2}=\epsilon^{0}(\bar k_1\bar k_14-\bar k_1\bar x_2)\\
    \dot{\bar x}_{4}=\epsilon^{-2}(\bar k_4\bar x_2\bar x_5 + 2\bar k_9\bar k_{14}\bar x_2\bar x_4- \\
    -\bar k_9k_{14}^2\bar x_4 - \bar k_9\bar x_2^2\bar x_4)\\
    \dot{\bar x}_{5}=\epsilon^{-2}(\bar k_6-\bar k_4\bar x_2\bar x_5)
\end{array}$  
&
$\begin{array}{l}
x_3 \leftarrow k_{14}-x_1-x_2-x_4 \\
x_2 \leftarrow x_1+x_2
\end{array}$
 & 
 $\begin{array}{l}
 \dot x_{2}=k_1k_{14} - k_1x_2\\
    x_1=(k_3x_2)/k_2\\
    x_4=k_6/(k_9k_{14}^2 + k_9x_2^2 -\\
    -2k_9k_{14}x_2)\\
    x_5=k_6/(k_4x_2)
 \end{array}$    
\\
\hline
 $\begin{array}{l}  \{x_3,x_4\}_{05} \\(3,     0,     0,     3,     4)\end{array}$ & 
$\begin{array}{l}
    \dot{\bar x}_{1}=\epsilon^{-6}(\bar k_3\bar x_2-\bar k_2\bar x_1)\\
    \dot{\bar x}_{2}=\epsilon^{0}(\bar k_1\bar k_14-\bar k_1\bar x_2)\\
     \dot{\bar x}_{4}= \epsilon^{-2}(2\bar k_9 \bar k_{14} \bar x_2\bar x_4-\\
     -\bar k_9\bar k_{14}^2\bar x_4 
     - \bar k_9\bar x_2^2\bar x_4)\\
    \dot{\bar x}_{5}=\epsilon^{-2}(\bar k_6-\bar k_4\bar x_2\bar x_5)
\end{array}$  
&
$\begin{array}{l}
x_3 \leftarrow k_{14}-x_1-x_2-x_4 \\
x_2 \leftarrow x_1+x_2
\end{array}$
 & 
 $\begin{array}{l}
    \dot x_{2}=k_1k_14 - k_1x_2\\
    \dot x_{4}=2k_9k_14x_2x_4 - k_9x_2^2x_4 -\\- k_9k_{14}^2x_4\\
    \dot x_{5}=k_6 - k_4x_2x_5\\
    x_1=(k_3x_2)/k_2
 \end{array}$    
\\
\hline
$\begin{array}{l}  \{x_3,x_4\}_{04} \\( 3,     0,     1,     3,    4)\end{array}$ & 
$\begin{array}{l}
    \dot{\bar x}_{1}=\epsilon^{-6}(\bar k_3\bar x_2-\bar k_2\bar x_1)\\
    \dot{\bar x}_{2}=\epsilon^{1}(\bar k_1\bar x_3)\\
    \dot{\bar x}_{3}=\epsilon^{0}(-\bar k_1\bar x_3)\\
    \dot{\bar x}_{4}=\epsilon^{-1}(\bar k_4\bar x_2\bar x_5)\\
    \dot{\bar x}_{5}=\epsilon^{-2}(\bar k_6-\bar k_4\bar x_2\bar x_5)
\end{array}$  
&
$x_2 \leftarrow x_1+x_2$
 & 
 $\begin{array}{l}
    \dot x_{2}=k_1x_3\\
    \dot x_{3}=-k_1x_3\\
    \dot x_{4}=k_6\\
    x_1=(k_3x_2)/k_2\\
    x_5=k_6/(k_4x_2)
 \end{array}$    
\\
\hline
$\begin{array}{l}  \{x_3\}_{00} \\( 3,     0,     1,     2,     4)\end{array}$ & 
$\begin{array}{l}
    \dot{\bar x}_{1}=\epsilon^{-6}(\bar k_3\bar x_2-\bar k_2\bar x_1)\\
    \dot{\bar x}_{2}=\epsilon^{1}(\bar k_1\bar x_3)\\
    \dot{\bar x}_{3}=\epsilon^{0}(-\bar k_1\bar x_3)\\
    \dot{\bar x}_{4}=\epsilon^{0}(\bar k_4\bar x_2\bar x_5-\bar k_9\bar x_3^2\bar x_4)\\
    \dot{\bar x}_{5}=\epsilon^{-2}(\bar k_6-\bar k_4\bar x_2\bar x_5)
\end{array}$  
&
$x_2 \leftarrow x_1+x_2$
 & 
 $\begin{array}{l}
    \dot x_{2}=k_1x_3\\
    \dot x_{3}=-k_1x_3\\
    \dot x_{4}=k_6 - k_9x_3^2x_4\\
    x_1=(k_3x_2)/k_2\\
    x_5=k_6/(k_4x_2)
 \end{array}$    
\\
\hline
$\begin{array}{l}  \{x_3,x_4\}_{04} \\( 3,     0,     2,     2,     4)\end{array}$ & 
$\begin{array}{l}
    \dot{\bar x}_{1}=\epsilon^{-6}(\bar k_3\bar x_2-\bar k_2\bar x_1)\\
    \dot{\bar x}_{2}=\epsilon^{2}(\bar k_1\bar x_3-\bar k_4\bar x_2\bar x_5)\\
    \dot{\bar x}_{3}=\epsilon^{0}(-\bar k_1\bar x_3)\\
    \dot{\bar x}_{4}=\epsilon^{0}(\bar k_4\bar x_2\bar x_5)\\
    \dot{\bar x}_{5}=\epsilon^{-2}(\bar k_6-\bar k_4\bar x_2\bar x_5)
\end{array}$  
&
$x_2 \leftarrow x_1+x_2$
 & 
 $\begin{array}{l}
    \dot x_{2}=k_1x_3 - k_6\\
    \dot x_{3}=-k_1x_3\\
    \dot x_{4}=k_6\\
    x_1=(k_3x_2)/k_2\\
    x_5=k_6/(k_4x_2)
 \end{array}$    
\\
\hline
$\begin{array}{l}  \{x_3,x_4\}_{04} \\( 3,     0,     2,     1,     4)\end{array}$ & 
$\begin{array}{l}
    \dot{\bar x}_{1}=\epsilon^{-6}(\bar k_3\bar x_2-\bar k_2\bar x_1)\\
    \dot{\bar x}_{2}=\epsilon^{2}(\bar k_1\bar x_3-\bar k_4\bar x_2\bar x_5)\\
    \dot{\bar x}_{3}=\epsilon^{0}(-\bar k_1\bar x_3)\\
    \dot{\bar x}_{4}=\epsilon^{1}(\bar k_4\bar x_2\bar x_5)\\
    \dot{\bar x}_{5}=\epsilon^{-2}(\bar k_6-\bar k_4\bar x_2\bar x_5)
\end{array}$  
&
$x_2 \leftarrow x_1+x_2$
 & 
 $\begin{array}{l}
    \dot x_{2}=k_1x_3 - k_6\\
    \dot x_{3}=-k_1x_3\\
    \dot x_{4}=k_6\\
    x_1=(k_3x_2)/k_2\\
    x_5=k_6/(k_4x_2)
 \end{array}$    
\\
\hline
$\begin{array}{l} \{x_4\}_{00} \\( 3,     0,     3,     1,     4) \end{array}$ & 
$\begin{array}{l}
    \dot{\bar x}_{1}=\epsilon^{-6}(\bar k_3\bar x_2-\bar k_2\bar x_1)\\
    \dot{\bar x}_{2}=\epsilon^{2}(-\bar k_4\bar x_2\bar x_5)\\
    \dot{\bar x}_{3}=\epsilon^{0}(\bar k_{10}\bar x_4-\bar k_1\bar x_3)\\
    \dot{\bar x}_{4}=\epsilon^{1}(\bar k_4\bar x_2\bar x_5)\\
    \dot{\bar x}_{5}=\epsilon^{-2}(\bar k_6-\bar k_4\bar x_2\bar x_5)
\end{array}$  
&
$x_2 \leftarrow x_1+x_2$
 & 
 $\begin{array}{l}
    \dot x_{2}=-k_6\\
    \dot x_{3}=k_{10}x_4 - k_1x_3\\
    \dot x_{4}=k_6\\
    x_1=(k_3x_2)/k_2\\
    x_5=k_6/(k_4x_2)
 \end{array}$    
\\
\hline
$\begin{array}{l}  \{\emptyset\}_{00} \\ ( 3,     0,     2,     0,     4) \end{array}$  & 
$\begin{array}{l}
    \dot{\bar x}_{1}=\epsilon^{-6}(\bar k_3\bar x_2-\bar k_2\bar x_1)\\
    \dot{\bar x}_{2}=\epsilon^{2}(\bar k_1\bar x_3-\bar k_4\bar x_2\bar x_5)\\
    \dot{\bar x}_{3}=\epsilon^{0}(\bar k_{10}\bar x_4 + \bar k_9\bar x_3^2\bar x_4-\bar k_1\bar x_3)\\
    \dot{\bar x}_{4}=\epsilon^{2}(\bar k_4\bar x_2\bar x_5- \bar k_{10}\bar x_4 - \bar k_9\bar x_3^2\bar x_4)\\
    \dot{\bar x}_{5}=\epsilon^{-2}(\bar k_6-\bar k_4\bar x_2\bar x_5)\\
\end{array}$  
&
$x_2 \leftarrow x_1+x_2$
 & 
 $\begin{array}{l}
    \dot x_{2}=k_1x_3 - k_6\\
    \dot x_{3}=k_{10}x_4 - k_1x_3 + k_9x_3^2x_4\\
    \dot x_{4}=k_6 - k_{10}x_4 - k_9x_3^2x_4\\
    x_1=(k_3x_2)/k_2\\
    x_5=k_6/(k_4x_2)
 \end{array}$    
\\
\hline
$\begin{array}{l} \{x_4\}_{02} \\( 3,     0,     1,     1,     4)\end{array}$ & 
$\begin{array}{l}
    \dot{\bar x}_{1}=\epsilon^{-6}(\bar k_3\bar x_2-\bar k_2\bar x_1)\\
    \dot{\bar x}_{2}=\epsilon^{1}(\bar k_1\bar x_3)\\
    \dot{\bar x}_{3}=\epsilon^{0}(\bar k_9\bar x_3^2\bar x_4-\bar k_1\bar x_3)\\
    \dot{\bar x}_{4}=\epsilon^{0}(-\bar k_9\bar x_3^2\bar x_4)\\
    \dot{\bar x}_{5}=\epsilon^{-2}(\bar k_6-\bar k_4\bar x_2\bar x_5)
\end{array}$  
&
$x_2 \leftarrow x_1+x_2$
 & 
 $\begin{array}{l}
    \dot x_{2}=k_1x_3\\
    \dot x_{3}=k_9x_3^2x_4 - k_1x_3\\
    \dot x_{4}=-k_9x_3^2x_4\\
    x_1=(k_3x_2)/k_2\\
    x_5=k_6/(k_4x_2)
 \end{array}$    
\\
\hline
\end{longtable}
 
}

\end{center}

\end{document}